\documentclass[12pt]{article}
\usepackage[pdftex]{graphicx}
\usepackage{natbib}
\usepackage[utf8]{inputenc}

\usepackage{multirow}
\usepackage{hyperref}
\usepackage{indentfirst}
\usepackage{amsfonts, amssymb, amsmath, amsthm, bm} 
\usepackage{dsfont, accents}

\usepackage{float, enumerate}
\usepackage[mathscr]{euscript}

\usepackage{algorithm}
\usepackage{algpseudocode}
\usepackage{float}

\usepackage{caption}
\usepackage{subcaption}
\usepackage[section]{placeins}
\usepackage{authblk}

\title{Market Simulation under Adverse Selection}
\author[1]{Luca Lalor}
\author[1]{Anatoliy Swishchuk}
\affil[1]{\small University of Calgary, Department of Mathematics and Statistics, Calgary, AB T2N 1N4, Canada}
\date{\today}

\begin{document}
\maketitle

\thispagestyle{empty}

\begin{abstract}
In this paper, we study the effects of fill probabilities and adverse fills on the trading strategy simulation process.  We specifically focus on a stochastic optimal control market-making problem and test the strategy on ES (E-mini S\&P 500), NQ (E-mini Nasdaq 100), CL (Crude Oil) and ZN (10-Year Treasury Note), which are some of the most liquid futures contracts listed on the CME (Chicago Mercantile Exchange). We provide empirical evidence that shows how fill probabilities and adverse fills can significantly affect performance and propose a more prudent simulation framework to deal with this. Many previous works aim to measure different types of adverse selection in the limit order book (LOB), however, they often simulate price processes and market orders independently. This has the ability to largely inflate the performance of a short-term style trading strategy. Our studies show that using more realistic fill probabilities and tracking adverse fills in the strategy simulation process more accurately shows how these types of trading strategies would perform in reality. 
\end{abstract}


\pagenumbering{arabic}

\newpage
\section{Introduction}

Over the past two decades, algorithmic trading and high-frequency trading (HFT) have become the dominant mechanisms for executing transactions in the world’s leading financial markets. Comprehensive surveys by \cite{gould2013limit} and \cite{jain2024limit} review both the empirical and theoretical literature on limit order books and HFT, while also highlighting several important shortcomings in existing modeling approaches. 
Trading algorithms can submit many different order types, but this paper focuses exclusively on market orders (MOs)\footnote{ Market Orders execute immediately at the best available price.} and limit orders (LOs)\footnote{Limit Orders rest in the limit order book (LOB) until they are either executed against an incoming MO or canceled.}, as is common in much of the literature since most other order types (along with trade order cancellations) can be viewed as extensions or combinations of these basic mechanisms. 




Stochastic processes have long played a central role in modeling algorithmic and high-frequency trading markets. Works such as \cite{cartea2014buy}, \cite{cartea2015algorithmic}, \cite{bulthuis2017optimal}, \cite{cartea2018enhancing}, \cite{cartea2018algorithmic}, and \cite{roldan2022optimal} present a broad range of Stochastic Optimal Control (SOC) frameworks for problems including optimal liquidation and acquisition, market making, volume-targeted trading, pairs trading, statistical arbitrage, order imbalance, price impact modeling, and Hawkes process-based limit order book event dynamics. However, while such predictive models may offer advantages in some scenarios, they cannot fully account for random, HFT order flow, where sudden spikes in activity occur too quickly for an effective response.

The type of problem that this paper will focus on is the Market Maker (MM) style problem given in Section 10.4.2 of \cite{cartea2015algorithmic}, where the stochastic control is whether or not the MM should post one unit (be it shares/lots etc) on the best bid/ask. This problem was chosen because it highlights two of the major issues for an MM that are often overlooked and oversimplified in the literature, the fill probability and the adverse fill. These can be summarized as follows:

\begin{itemize}
\item Fill Probability: The probability that a trader’s LO is executed after being posted in the Limit Order Book (LOB). This depends on factors such as the queue position, available liquidity at the price level, and the arrival rate and size of the incoming market orders. A common proxy for market depth is the size of the bid and ask queues in the LOB, while a more heuristic alternative considers the ratio of traded volume to resting limit order sizes. As an example, see Table \ref{tab:tabletradesize} which reports the mean and median trade sizes per contract for four of the most liquid U.S. futures contracts, computed from LOB data on April 9, 2024.

\item Adverse fills: A form of adverse selection that occurs when a passive MMs LO is executed at a disadvantageous price. Specifically, this happens when the MM’s order is ``picked off," which means that immediately after execution, the new trade position is out of the money when marked-to-market. We will demonstrate that adverse fills occur in the majority of trade executions across various assets. Based on this empirical evidence, we develop a method to incorporate adverse fills into our proposed trading simulation environment, ensuring a more realistic representation of execution risk.
\end{itemize}
\begin{table}[h!]
  \begin{center}
    \begin{tabular}{l|c|r} 
      \textbf{Contract} & \textbf{Mean Trade Size} & \textbf{Median Trade Size} \\
      \hline
      ES & 3.47 & 1\\
      NQ & 1.54 & 1\\
      CL & 1.70 & 1\\
      ZN & 16.01 & 2
    \end{tabular}
    \captionsetup{font=small}
    \caption{The mean and median trade sizes in four of the most liquid Chicago Mercantile Exchange (CME) futures contracts.}
    \label{tab:tabletradesize}
  \end{center}
\end{table}

Although some SOC frameworks incorporate adverse fills, prior work has not explicitly modified SOC trading dynamics or LOB models to distinguish between adverse and non-adverse executions. In practice, adverse fills cannot be fully eliminated, regardless of model sophistication. Motivating the need to treat these outcomes separately within an SOC setting, similar findings to ours also appear in \cite{law2019market} and \cite{delise2024negativedrift}, where the former highlights how diffusion-based MM models can create ``phantom gains", and the latter provides an empirical analysis of the ``negative drift of a limit order fill", a key factor in an MM's performance. 

Subsequently, the main contributions of this paper can be summarized as follows:
\begin{itemize}
    \item We develop an enhanced SOC simulation framework for high-frequency trading strategies in a standard optimal MM setting. The framework explicitly models both adverse and non-adverse fills, relaxing the common assumption of independence between trade executions and price dynamics. We show that incorporating this distinction can materially affect strategy performance and significantly alters evaluation outcomes.
    \item We introduce a probability measure specifically for non-adverse fills to refine their modeling within MM strategies. As shown in Section 3, naive strategies are exposed to excessive adverse fills and do not guarantee non-adverse execution for every incoming market order. While fill probabilities are commonly used in the literature, we argue they should be applied only to non-adverse fills, since adverse fills arise mechanically from limit order book (LOB) execution rules. This distinction yields performance measures that better reflect real-world trading conditions.
\end{itemize}

The rest of this paper will proceed as follows. In Section 2, we begin by presenting two simple motivating examples of a MM trading strategy that illustrate the different types of trade order fills, together with empirical evidence on how frequently they may occur. In Section 3, we introduce an SOC MM problem, which we will use for our simulation analysis. In Section 4, we explain how one can create a simulation environment by discretizing the continuous processes given in Section 3. Here we compare the simulation environment in \cite{cartea2015algorithmic} and \cite{jaimungalgit}, which we will refer to as the benchmark environment, with our improved simulation environment. Then, in Section 5, we show the performance results of the trading strategy in both simulation environments. Lastly, this paper ends with some concluding remarks and ideas for future research.

\section{Adverse and Non-adverse Trade Order Fills}

Most existing simulation models in the literature assume independence between price dynamics and market order arrivals, which can lead to inflated performance metrics. In particular, adverse fills are often ignored or mischaracterized, while ad hoc probabilities are assigned to non-adverse executions. This is problematic because, in a LOB, adverse fills are mechanically induced: if a limit order is posted at a given price and the asset price moves through that level, execution is effectively guaranteed at a worse price. Ignoring this deterministic feature leads to an inconsistent representation of the fill process and distorts the evaluation of trading strategies, often inflating performance measures.

To illustrate this, consider a real-world LOB sampled at 1-second intervals, as used in our empirical analysis. Empirical evidence shows that, for most assets, best bid and ask prices remain unchanged far more frequently than they move at each time step, as documented in \cite{biais1995empirical}, \cite{hasbrouck2013low}, and \cite{cont2014price}, among others.
When MOs are modeled independently of price dynamics, executions are not necessarily aligned with price crossings. Unless fills are explicitly enforced when price levels are breached, simulated executions will rarely capture adverse fills, leading to a systematic underestimation of their frequency. This issue becomes more pronounced at higher sampling frequencies: as time steps shrink and price changes become rarer within each interval, the likelihood of correctly aligning fills with price movements decreases further. In Section 4 we provide empirical evidence showing that neglecting adverse fills can substantially distort strategy performance evaluation.

In the remainder of this section, we present two examples to illustrate key concepts and results. Section 2.1 introduces a simple example to clarify the distinction between adverse and non-adverse fills, serving purely as an illustration of the two trade-order fill types. Section 2.2 explores a basic LO posting strategy for an MM, simulated in a real-time trading environment. This second example provides insight into the expected distribution of adverse and non-adverse fills for a simple MM strategy, particularly in the context of relatively small order sizes.

\subsection{Example 1: Simple Market Maker Trade Order Fills}

We consider a simple LOB MM setting to illustrate the distinction between adverse and non-adverse fills. The model abstracts from strategic behavior and focuses purely on execution dynamics. This example isolates execution mechanics from strategic decision-making to highlight how fills are classified in a dynamic LOB environment. In summary, at each time step:
\begin{itemize}
    \item The best bid and ask evolve as a random walk with a fixed spread of one cent.
    \item The MM agent posts a unit-sized LO at both the best bid and best ask.
    \item One MO arrives and randomly executes against either the bid or ask side.
    \item The MM agent is assumed to be at the front of the queue, so every incoming MO fully fills the posted LO (i.e., 100\% fill probability).
\end{itemize}

\begin{figure}[H]
	\begin{center}
		\includegraphics[width=\linewidth]{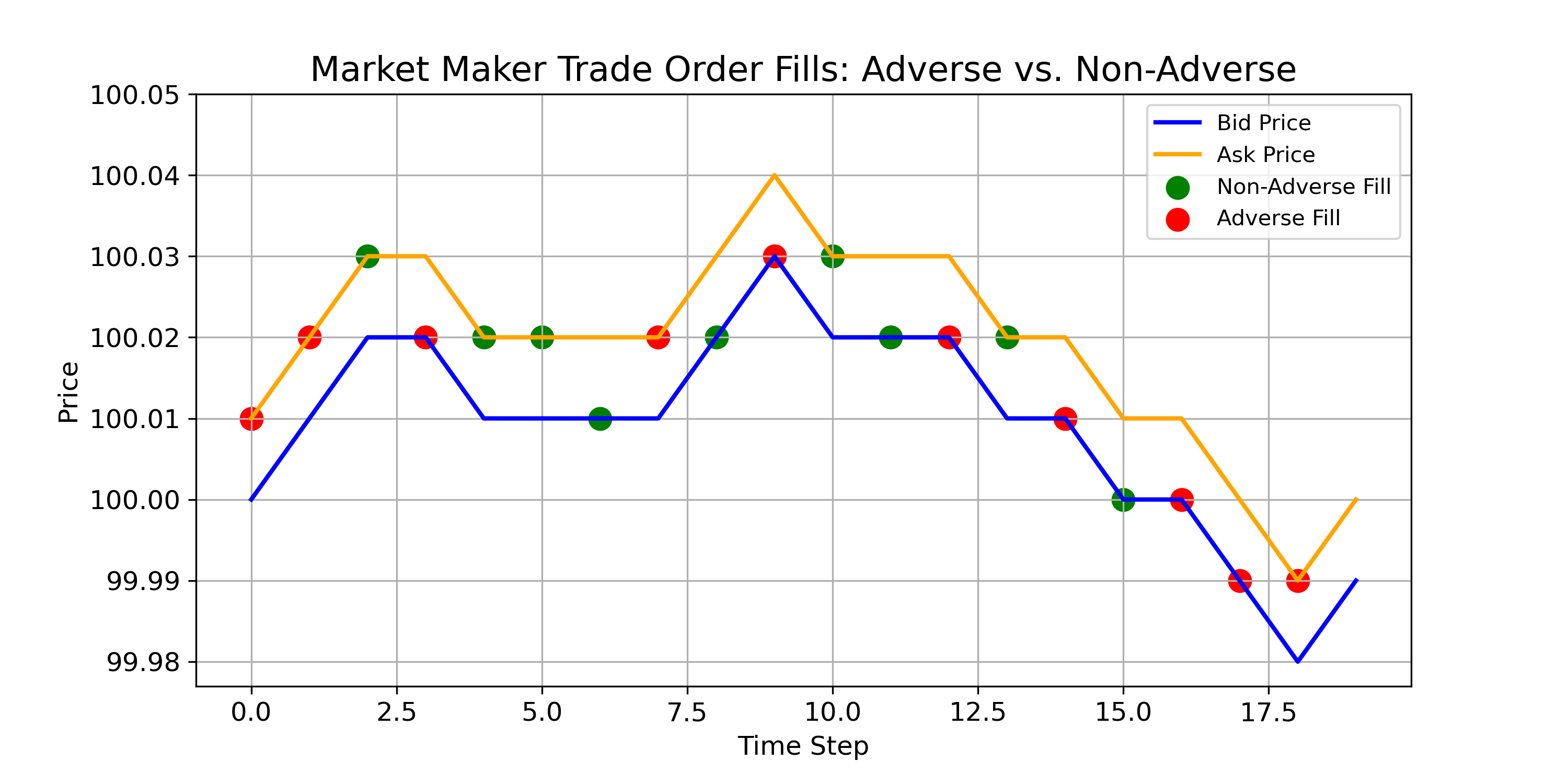}
		\captionsetup{font=small}
  		\caption{A visualization of the simple MM strategy in Example 1, highlighting the timing of each trade order fill and distinguishing between the different trade order fill types.}
  		\label{fig:Ex_1_SimpleMM}
	\end{center}
\end{figure}

Figure \ref{fig:Ex_1_SimpleMM} illustrates the evolution of the best bid (blue) and best ask (yellow) prices over time. Red markers indicate adverse fills, while green markers indicate non-adverse fills, clearly distinguishing between the two execution outcomes. The figure highlights that, under this execution assumption, price movements against the MM are immediately associated with executions at the prevailing quotes. As a result, adverse selection directly impacts the marked-to-market value of the inventory at the time of the fill.

\subsection{Example 2: Market Maker Trade Order Fills in a Real-Time Simulation Environment}

To further illustrate the importance of distinguishing between adverse and non-adverse fills, we analyze LO execution data from the Trading Technologies (TT) simulation environment. TT is a professional trading platform that provides live market access and a realistic simulator to test strategies without risking capital. For relatively small order sizes, simulated performance can often match real-market behavior. Using simulated trades over a full trading day (8:30 EST--16:00 EST), we examine the frequency of adverse and non-adverse fills. TT models LO executions by assigning orders an imaginary queue position in the LOB, which is filled once sufficient real market volume trades through that position. This produces a realistic approximation of fill dynamics, particularly when order sizes are too small to materially impact prices.

To show the likelihood of adverse fills occurring, we start by simulating a simple trading strategy in the following near-term expiry futures contracts: E-mini S\&P 500 (ES), Nasdaq 100 (NQ), WTI Crude Oil (CL) and 10-Year Treasury Note (ZN), which posts LOs close to the best bid/ask over a single trading day. The main components of the strategy can be described as follows. For ES and CL, the strategy posted static 1-lot bid and ask orders 4 ticks apart, while the spacing was 16 ticks for NQ and 1 tick for ZN. These distances were chosen to generate relatively low trading volume while still producing enough executions to analyze empirical LO fill behavior. The differing tick spacings reflect variations in trading activity and price dynamics across assets. For example, NQ exhibits substantially more frequent price changes than ES, CL, and especially ZN, which often remains unchanged for extended periods. Selecting the order spacing in this manner produced a more balanced and comparable dataset across assets. The strategy also reposts orders at the same filled price whenever a fill occurs at a different price. See Algorithm \ref{alg:StaticRepostMM} below for a formal description, where $S^{(a)}$, $S^{(b)}$, and $\Delta$ represent the ask price, bid price, and quote spacing, respectively.

\begin{algorithm}
\caption{Static Reposting Market-Making Strategy}
\label{alg:StaticRepostMM}
\begin{algorithmic}[1]
\State Initialize bid LO at price $S_0^{(b)}$
\State Initialize ask LO at price $S_0^{(a)}$
\State Let $\Delta > 0$ denote the fixed quote spacing
\State Let $t$ denote non-uniform event-driven time steps
\For{each event time $t$}
    \If{bid LO at price $S^{(b)}$ is filled}
        \State Post new bid LO at $S^{(b)} - \Delta$
        \State Post ask LO at $S^{(b)} + \Delta$ if not already active
    \EndIf
    \If{ask LO at price $S^{(a)}$ is filled}
        \State Post new ask LO at $S^{(a)} + \Delta$
        \State Post bid LO at $S^{(a)} - \Delta$ if not already active
    \EndIf
    \State Previously posted outer orders remain active
\EndFor
\end{algorithmic}
\end{algorithm}


As in example 1, each trade order fill that precedes a disadvantageous move in the price will be considered adverse, while every other fill will be considered non-adverse. More specifically, when an LO is filled on the bid (ask) and the next new bid (ask) price is lower (higher), we consider this fill to be adverse, as the first change in the marked-to-market value of the position is negative. 
See Table \ref{tab:tablemotivex} for the results, where we can see that a significant portion of the total number of LO fills in ES, NQ, CL and ZN were adverse. 

\begin{table}[h!]
	\begin{center}
		\begin{tabular}{ |p{2cm}|p{2cm}|p{2cm}|p{2.5cm}|p{2.5cm}|  }
 		\hline
 		\multicolumn{5}{|c|}{\textbf{Trade Order Fills}} \\
 		\hline
 	\textbf{Date}& \textbf{Contract} & \textbf{Total number of LO fills} 		&\textbf{Number of adverse fills} & \textbf{Number of non-adverse fills}\\
 		\hline
 		2024/04/24 & ES Jun24 & 941 & 767 & 174\\
      	2024/04/25 & NQ Jun24 & 1929 & 1269 & 660 \\
      	2024/04/23 & CL Jun24 & 625 & 518 & 107 \\
      	2024/04/24 & ZN Jun24 & 224 & 199 & 25\\

 		\hline
		\end{tabular}
	\captionsetup{font=small}
	\caption{The different types of LO fills for the basic posting strategy over active trading trading hours (8:30 EST - 16:00 EST). }
	\label{tab:tablemotivex}
	\end{center}
\end{table}

These empirical results demonstrate that adverse fills can materially affect the performance of short-term intraday trading strategies that post LOs, regardless of the underlying strategy. This motivates the need for simulation frameworks that explicitly account for adverse fills, particularly in MM settings. In Section 3, we consider a simple SOC MM strategy that is unlikely to predict future order flow or price movements as accurately as professional market makers using proprietary models. Nevertheless, such strategies remain common benchmark models in the academic literature and teaching environment. More sophisticated MM strategies are difficult to replicate in practice because they are highly confidential and subject to alpha decay once publicly disclosed; see \cite{zhou2017alpha} and \cite{sivaramakrishnan2018multi}. Despite this, execution performance measures are largely strategy independent, implying that similar adverse and non-adverse fill dynamics should still arise in our SOC MM framework. Subsequently, we develop an improved MM simulation environment, showing that results change materially once the model explicitly accounts for adverse fills and incorporates realistic non-adverse fill probabilities. We focus on MM strategies because they often trade passively, posting large volumes of LOs in the LOB and executing substantial trade flow throughout the day.

\section{A Stochastic Optimal Control Market-Making Problem}

An MM problem in the SOC framework involves a trader seeking to maximize terminal wealth while trading a large quantity of an asset, typically via limit orders. In our example, the control variable is whether to post at the best bid or ask, a standard MM decision. The key feature is that the problem introduces adverse selection through a predictive signal, but fails to properly model order execution and fill dynamics, which is an important practical challenge in real-world market making.

Firstly, we would like to introduce the MM problem studied, as given in \citet{cartea2015algorithmic}, where we also introduce an additional probability measure for non-adverse fills. We begin by describing the key stochastic processes that satisfy certain SDEs as explained in \citet{cartea2015algorithmic} and \citet{cartea2018algorithmic} as follows:
\begin{itemize}
\item The midprice process satisfies the SDE,
\begin{align}
dS_t = (\nu+\alpha_t)dt+\sigma dW_t , 
\label{eq: mpcartea}
\end{align}
Here, the drift is given by a long-term component $\nu$ and by a short-term component $\alpha_t$, which is a predictable zero-mean reverting process, and $W_t$ is a standard arithmetic Brownian motion. 
\item The short-term drift term in Equation \eqref{eq: mpcartea}, $\alpha_t$, can be specified in several ways; following \citet{cartea2015algorithmic} and \citet{cartea2018algorithmic} for simplicity, we adopt their approach. The model assumes short-term alpha is driven by order flow, predicting price moves in the direction of incoming market orders: buy market orders generate upward price jumps, while sell market orders generate downward jumps. Consequently, $\alpha_t$ is modeled as a zero-mean-reverting process with random jump sizes at market order arrival times. Formally, $\alpha_t$ satisfies the SDE,
\begin{align}
d\alpha_t = -\zeta\alpha_tdt+\eta dW_t^{\alpha}+\epsilon_{1+M_t^-}^+dM_t^+ - \epsilon_{1+M_t^-}^-dM_t^-.
\label{eq: dalpcart}
\end{align}
Here, $\zeta$ and $\eta$ are positive constants, $W_t^{\alpha}$ is a Brownian motion independent of all other processes, $M_t^{\pm}$ is a counting processes for incoming MOs and ${\epsilon_1^{\pm}, \epsilon_2^{\pm},... }$ are i.i.d random variables also independent of all other processes representing the size of these MOs which arrive at an independent constant rate $\lambda^{\pm}$.
\item The controlled inventory process, denoted as $Q_t^{\delta}$, changes every time the agent's posted LO is filled by an incoming MO and can be stated as follows,
\begin{align} 
Q_t^{\delta}=N_t^{\delta,-}-N_t^{\delta, +},
\label{eq: invcartea}
\end{align}
where $N_t^{\delta,-}$ ($N_t^{\delta,+}$) are counting process that represent the buy (sell) LO fills that increase (decrease) the agents inventory. 
\item The agents cash process satisfies the SDE,
\begin{align}
dC_t^{\delta}=\displaystyle{\left(S_t+\frac{\Delta}{2}\right)}dN_t^{\delta, +}-\displaystyle{\left(S_t-\frac{\Delta}{2}\right)}dN_t^{\delta, -}.
\label{eq: cashcartea}
\end{align}
Here, $\Delta$ is the spread between the best bid and ask, and, as before, $N_t^{\delta, \pm}$ represents the counting process for filled LOs. 
\end{itemize}

The agents’ performance criteria, as in \citet{cartea2015algorithmic} and \citet{cartea2018algorithmic}, is then defined as follows,
\begin{align}
H^{\delta}(t,c,S,\alpha,q) = \mathbb{E}_{t,c,S,\alpha,q}\displaystyle{\left[C_T^{\delta}+Q_T^{\delta}\displaystyle{\left(S_T-\displaystyle{\left(\frac{\Delta}{2}+\varphi Q_T^{\delta}\right)}\right)}-\phi\int_t^T(Q_u^{\delta})^2du\right]},
\end{align}
where $\phi \int_t^TQ_u^2du$, with $\phi \ge{0}$, is a running inventory penalty which increases with time since the MMs goal is to have an inventory of zero at maturity $T$. The agents' value function is next defined as
\begin{align}
H(t,c,S,\alpha,q) = \sup_{\delta \in \mathscr{A}}H^{\delta}(t,c,S,\alpha,q),
\end{align}
where $\mathscr{A}$ is the set of admissible strategies in which $\delta >0$ and uniformly bounded from above. It is also important to note that there is an inventory constraint in this problem, by which the inventory process has an upper $\overline{q}$ and lower $\underline{q}$ bound. This essentially sets the maximum position that the MM can take long or short. Then, by applying the Dynamic Programming Principle (DPP), as explained in \citet{cartea2015algorithmic}, the value function will satisfy a long Dynamic Programming Equation (DPE), and then lead to an optimal control in feedback form as follows: 
\begin{align}
\begin{split}
\delta^{+,*}(t,\alpha,q)=\mathbb{I}_{\left\{ \frac{\Delta}{2}+\rho\mathbb{E}[h(t,\alpha+\epsilon^+,q-1)-h(t,\alpha+\epsilon^+,q)]>0 \right\} \cap \{q>\underline{q}\} }\\
\delta^{-,*}(t,\alpha,q)=\mathbb{I}_{\left\{ \frac{\Delta}{2}+\rho\mathbb{E}[h(t,\alpha-\epsilon^-,q+1)-h(t,\alpha-\epsilon^-,q)]>0 \right\} \cap \{q<\overline{q}\} }.\\
\end{split} \label{eq: optcontMM}
\end{align}

This optimal control problem closely follows the solution in \cite{cartea2015algorithmic}, where the full derivation can be found. A key difference is that their setup assumes a perfect fill probability and, therefore, omits execution uncertainty. Given the nature of this paper and the previous empirical evidence in Section 2, we relax this assumption by introducing a non-adverse fill probability $\rho(\delta^{\pm})$ in Equation \eqref{eq: optcontMM}, which we take to be either constant or independent of the state variables; for simplicity, we set $\rho_t^\delta = \rho \in (0,1)$. The non-adverse fill probability modification can be interpreted as follows:
\begin{itemize}
    \item If $0 < \rho < 1$, the posting threshold becomes more conservative, as the lower likelihood of execution reduces the expected benefit of placing a limit order.
    \item If $\rho = 1$, we recover the original case in \cite{cartea2015algorithmic}, corresponding to the benchmark environment, and the optimal controls in Equation \eqref{eq: optcontMM} remain unchanged.
\end{itemize}

The optimal decision in Equation \eqref{eq: optcontMM} then depends on the half-spread and the expected continuation value, evaluated at $\alpha \pm \epsilon^{\pm}$ and weighted by the fill probability. Intuitively, incoming MOs shift $\alpha$ upward or downward through random jumps, while execution occurs only with probability $\rho$. The resulting structure is almost identical to \cite{cartea2015algorithmic}, except for the introduction of this non-adverse fill probability. 

\section{Trading Simulation Environment}
In this section, we describe the trading simulation environment used to evaluate the optimal MM strategy. Following \citet{cartea2015algorithmic} and \citet{jaimungalgit}, we first implement their benchmark backtesting framework, originally based on simulated price dynamics. We extend this setup by replacing simulated prices with real LOB data, enabling later back-testing on actual futures contracts. We then introduce an improved simulation framework that more realistically captures MM dynamics, while still retaining some modeling limitations. In particular, we incorporate explicit tracking of adverse and non-adverse fills and refine the fill probability for non-adverse executions. We show that these extensions can have a substantial impact on strategy performance.

First, we briefly describe the simulation environment used in \citet{cartea2015algorithmic} and \citet{jaimungalgit} for the optimal market-making problem in Section 3, along with our extensions in more detail. The main components are as follows,
\begin{itemize}
\item Asset Price: We use futures LOB data sourced from Futures First Inc. As in Section 2, Example 2, we focus on near-term expiry contracts for ES, NQ, CL, and ZN. The data is resampled to 1-second intervals, which defines the time step in our discrete simulation. It includes up to five LOB levels, with each new market event updating at the next timestep. Because trades occur irregularly, resampling introduces missing intervals, which we handle using a forward-fill procedure that carries the most recent observation forward until a new update arrives, ensuring a continuous and consistent market state.
\item LO postings: Stores the decision/control variable, which tells us whether to be posted at the best bid/ask at any given time step as given by Equation \eqref{eq: optcontMM}.  
\item Trade order fills: The benchmark simulation environment from \cite{cartea2015algorithmic} and \citet{jaimungalgit} assumes all executions are non-adverse, except in rare cases where a fill coincides with an adverse price move. It also implicitly assumes that posted LOs are always at the front of the queue and, therefore, filled immediately upon arrival of a matching market order. In practice, especially in liquid markets, both assumptions are unrealistic. In our improved framework, we explicitly distinguish between adverse and non-adverse fills and introduce a fill probability $\rho$ (motivated by Section 2.2 and Tables \ref{tab:tabletradesize} and \ref{tab:tablemotivex}) governing non-adverse execution. Non-adverse fills at the ask and bid are defined as
\begin{align}
NFA_t &= \sum_{t_i \leq t} \delta^+_{t_i} \mathbb{I}_{\{M^+_{t_i}=1\}} \rho, \\
NFB_t &= \sum_{t_i \leq t} \delta^-_{t_i} \mathbb{I}_{\{M^-_{t_i}=1\}} \rho,
\label{eq: non-adv_fills}
\end{align}
where $NFA_t$ and $NFB_t$ count non-adverse executions at the best ask and bid up to time $t$. Here, $M^+_{t_i}$ and $M^-_{t_i}$ are counting processes that denote the arrival of buy and sell market orders, respectively, and $\rho$ is the probability that a posted order is executed as a non-adverse fill at time $t_i$. Unlike the benchmark case where $\rho=1$, in more general settings $\rho$ can be interpreted as capturing queue position, LOB dynamics, and market order intensity. However, since there is no well-established or robust method to calibrate this parameter that holds in real world live markets, we adopt a heuristic specification guided by the empirical evidence presented in Section 2.2. Next, adverse fills are tracked separately as
\begin{align}
AFA_{t} &= \sum_{t_i \leq t} \delta_{t_i}^+ \mathbb{I}_{\{S_{t_{i+1}}^{(a)} > S_{t_i}^{(a)}\}}, \\
AFB_{t} &= \sum_{t_i \leq t} \delta_{t_i}^- \mathbb{I}_{\{S_{t_{i+1}}^{(b)} < S_{t_i}^{(b)}\}},
\label{eq: adv_fills}
\end{align}
where $AFA_t$ and $AFB_t$ count adverse executions at the ask and bid, and $S_{t_i}^{(a)}$ and $S_{t_i}^{(b)}$ denote the ask and bid prices. Intuitively, if a LO is posted at the best bid (ask) and the next price move is downward (upward), the order is executed at the prevailing bid (ask) due to LOB priority rules, implying an adverse fill. Finally, total executed volume under the strategy is given by
\begin{align}
N_{t}^{\delta, +} &= \sum_{t_i \leq t} \left(AFA_{t_i} + NFA_{t_i}\right), \\
N_{t}^{\delta, -} &= \sum_{t_i \leq t} \left(AFB_{t_i} + NFB_{t_i}\right),
\label{eq: total_fills}
\end{align}
where $N_{t}^{\delta, +}$ and $N_{t}^{\delta, -}$ are counting processes that aggregate all ask- and bid-side fills up to time $t$ under the trading strategy.
\end{itemize}

It is also important to note that in \cite{cartea2015algorithmic} and \citet{jaimungalgit} the SOC problem is solved numerically and since a numerical solution is discrete, the values from this solution can be used directly to determine whether to be posted or not. For the processes involving MO arrivals, inventory, short-term alpha and cash, we follow the method given in \citet{jaimungalgit}, where they essentially discretize the Equations \eqref{eq: dalpcart}-\eqref{eq: cashcartea}, as well as devising a unique way to simulate MOs. 

\section{Simulation Results}

\begin{figure}[H]
    \centering
    \begin{subfigure}{0.45\textwidth}
        \includegraphics[width=\linewidth, height=5cm]{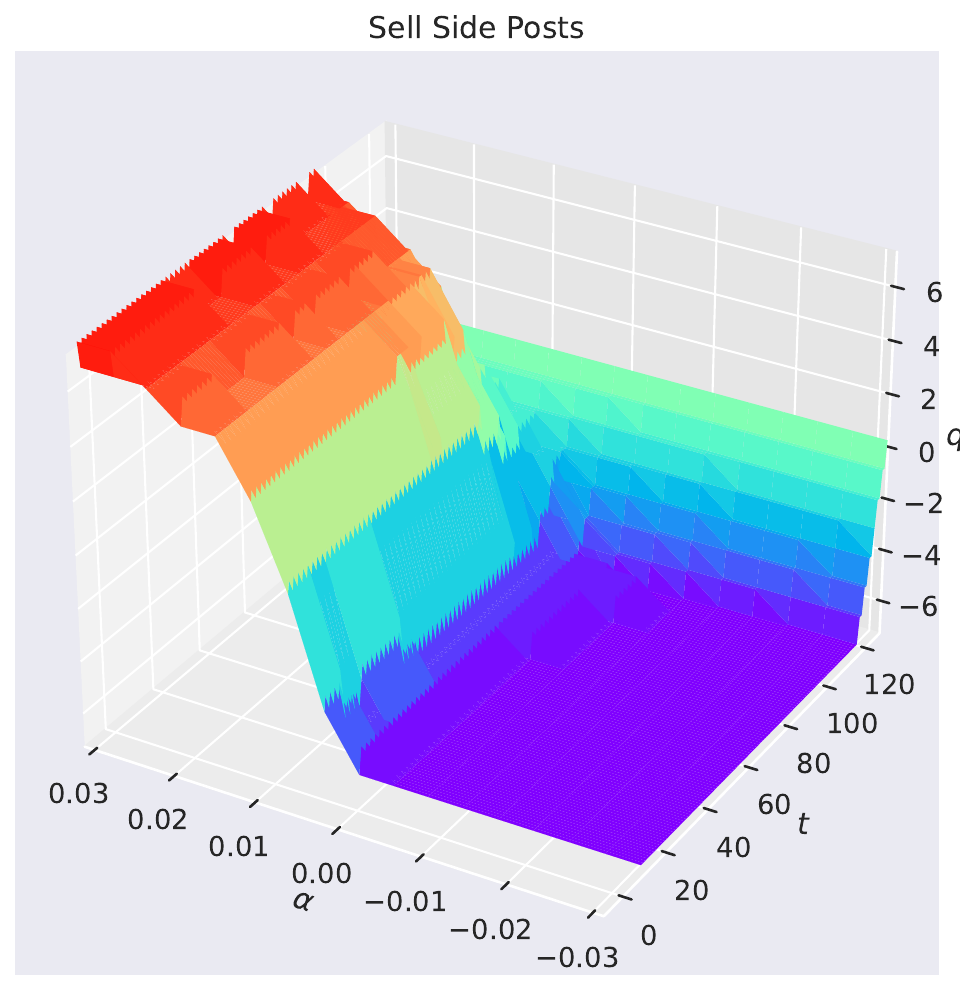} 
    \end{subfigure}
    \begin{subfigure}{0.45\textwidth}
        \includegraphics[width=\linewidth, height=5cm]{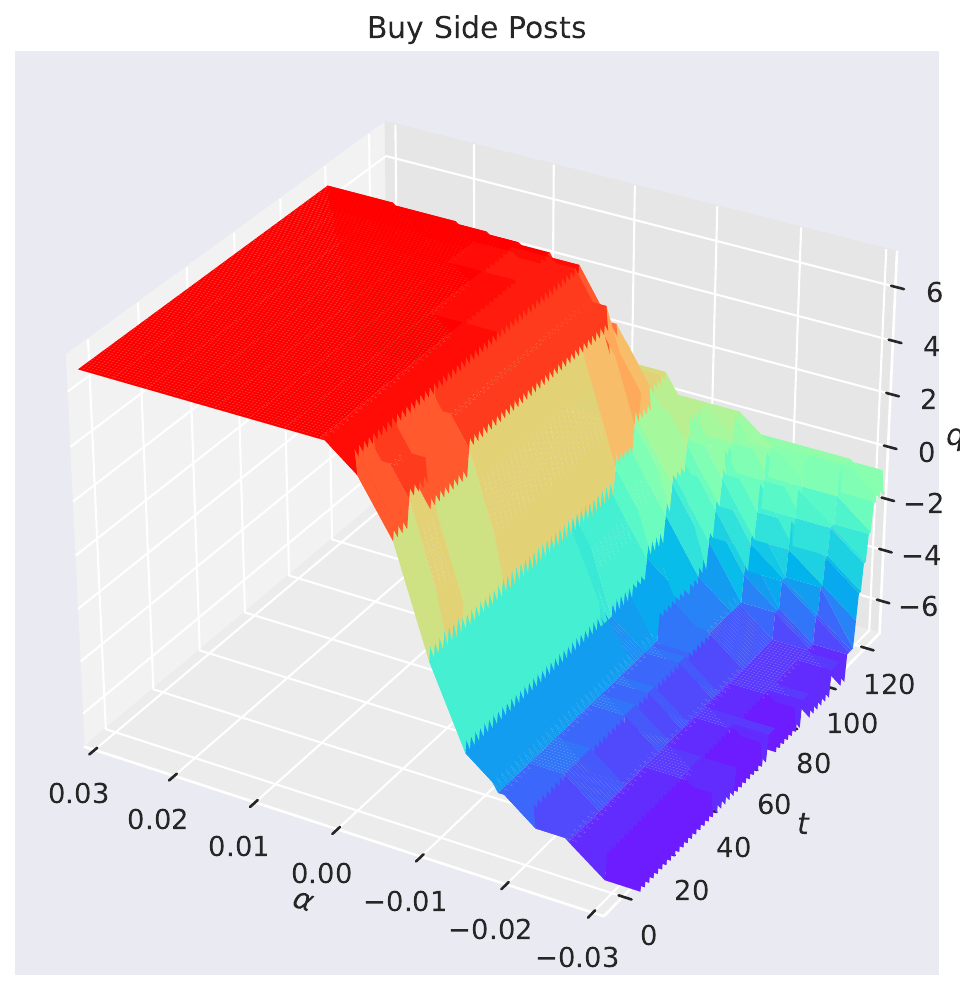}
    \end{subfigure}
    
    \begin{subfigure}{0.45\textwidth}
        \includegraphics[width=\linewidth, height=5cm]{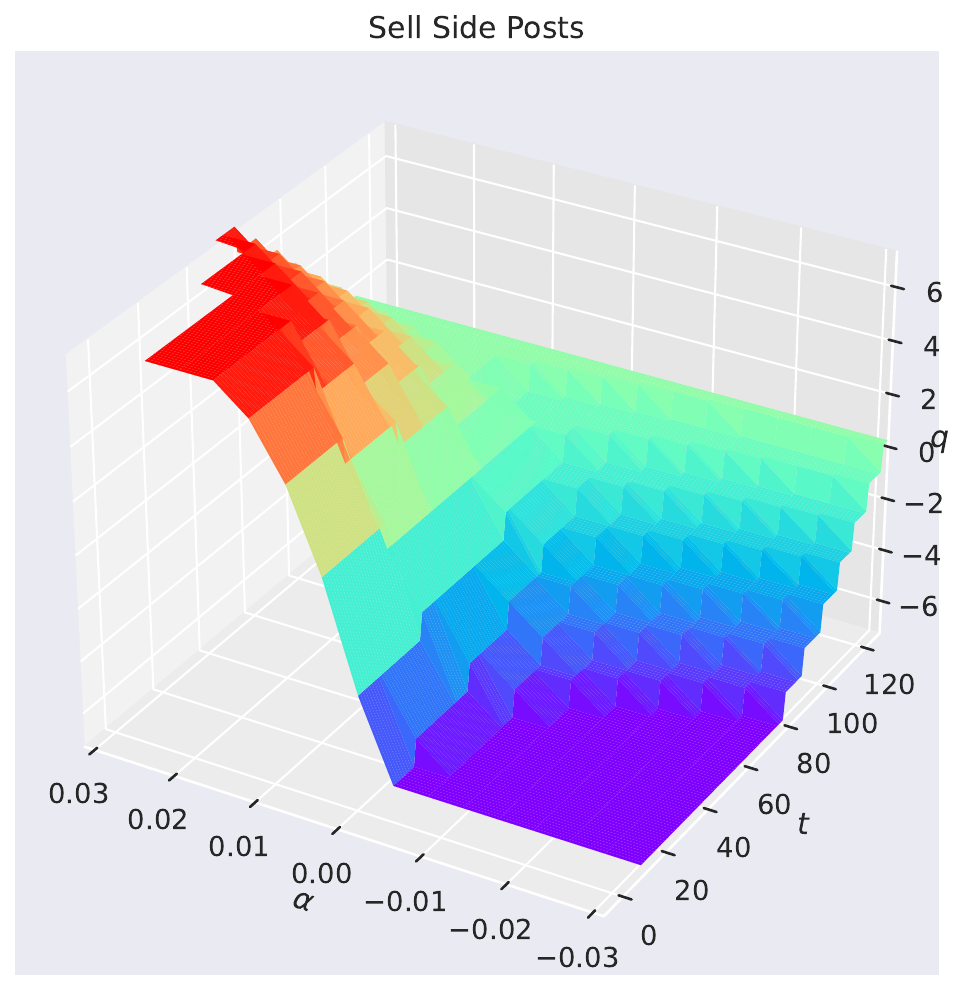}
    \end{subfigure}
    \begin{subfigure}{0.45\textwidth}
        \includegraphics[width=\linewidth, height=5cm]{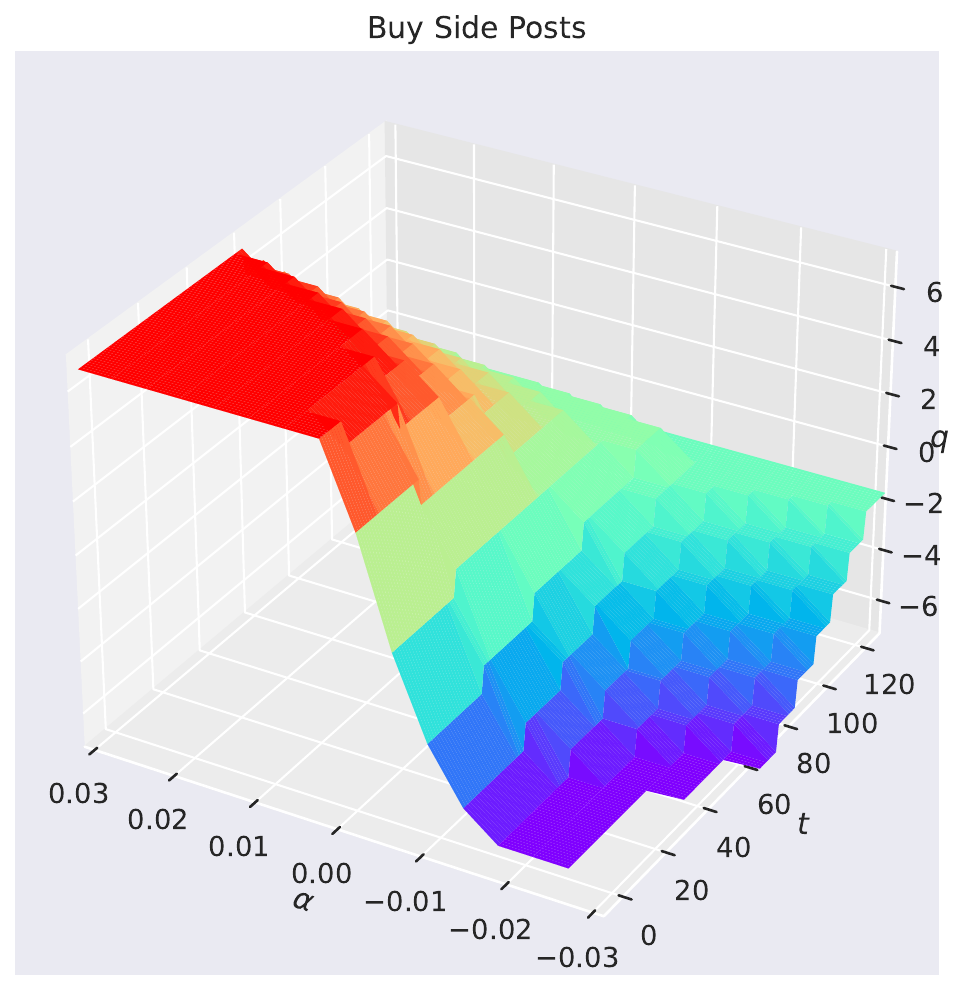}
    \end{subfigure}

    \captionsetup{font=small}
    \caption{Optimal solution as a function of time, short-term alpha, and inventory. The top plots correspond to $\rho = 1$, while the bottom plots correspond to $\rho = 0.2$. Here, darker red indicates a larger long position, while darker blue indicates a larger short position.}
    \label{fig:3Dplot}
\end{figure}

Here, we will discuss the results from testing the SOC MM strategy in our improved simulation environment and compare this with the benchmark environment. First, we would like to briefly summarize the numerical solution under our parameter set and provide a visual in Figure \ref{fig:3Dplot}, which looks similar to the solution given in \citet{cartea2015algorithmic} and \citet{jaimungalgit}. We compare the solutions of the buy and sell side posts, where the top plots correspond to $\rho=1$ and the bottom plots to $\rho=0.2$, and these values were chosen as these specific solutions will be utilized in our strategy simulations in Section 5. One can see how the optimal solution varies as a function of time ($t$), short-term alpha ($\alpha$) and inventory($q$). The left (right) figure displays a visual of LOs posted on the best offer (bid), where a sell (buy) LO is posted if the MMs inventory is above (below) the surface. It is clear that a lower non-adverse fill probability, $\rho$, leads to a slightly different solution where the agent is less likely to post LOs at lower inventory levels. The top left and top right plots, as well as the bottom left and bottom right plots, appear almost identical due to the equal arrival rates of MOs on both sides of the LOB. Far from maturity, the strategy is mostly independent of time, but one can see that as the MM approaches the maturity time ($T$), the optimal strategy is essentially independent of the short-term alpha ($\alpha$), since the MM is focused solely on liquidating any remaining position. 

As discussed earlier, the main difference in our improved simulation environment is that adverse fills are now included in the strategy simulation process, determined according to Equation \eqref{eq: adv_fills}.  Our aim in this section is to show how adverse fills and fill probabilities on non-adverse fills can significantly affect the performance of the trading strategy. We will discuss the results for CL in depth here, where the same repeated analysis for ES, NQ and ZN can be found in the appendix. Based on the empirical evidence for the distribution between adverse and non-adverse fills given in Table \ref{tab:tablemotivex}, we also set the non-adverse fill probability at $\rho=0.2$. See Table \ref{tab:fillssim}, below, for a list of the parameter values used. We kept most of the parameters similar to \citet{cartea2015algorithmic} and \citet{jaimungalgit}, however, we decreased the time step and increased the maturity time, to make the length of time more realistic to the data.  Some slight adjustments to some of the other parameters were also needed to keep the numerical solution stable. 

\begin{figure}[H]

\begin{subfigure}{0.5\textwidth}
\includegraphics[width=0.9\linewidth, height=6cm]{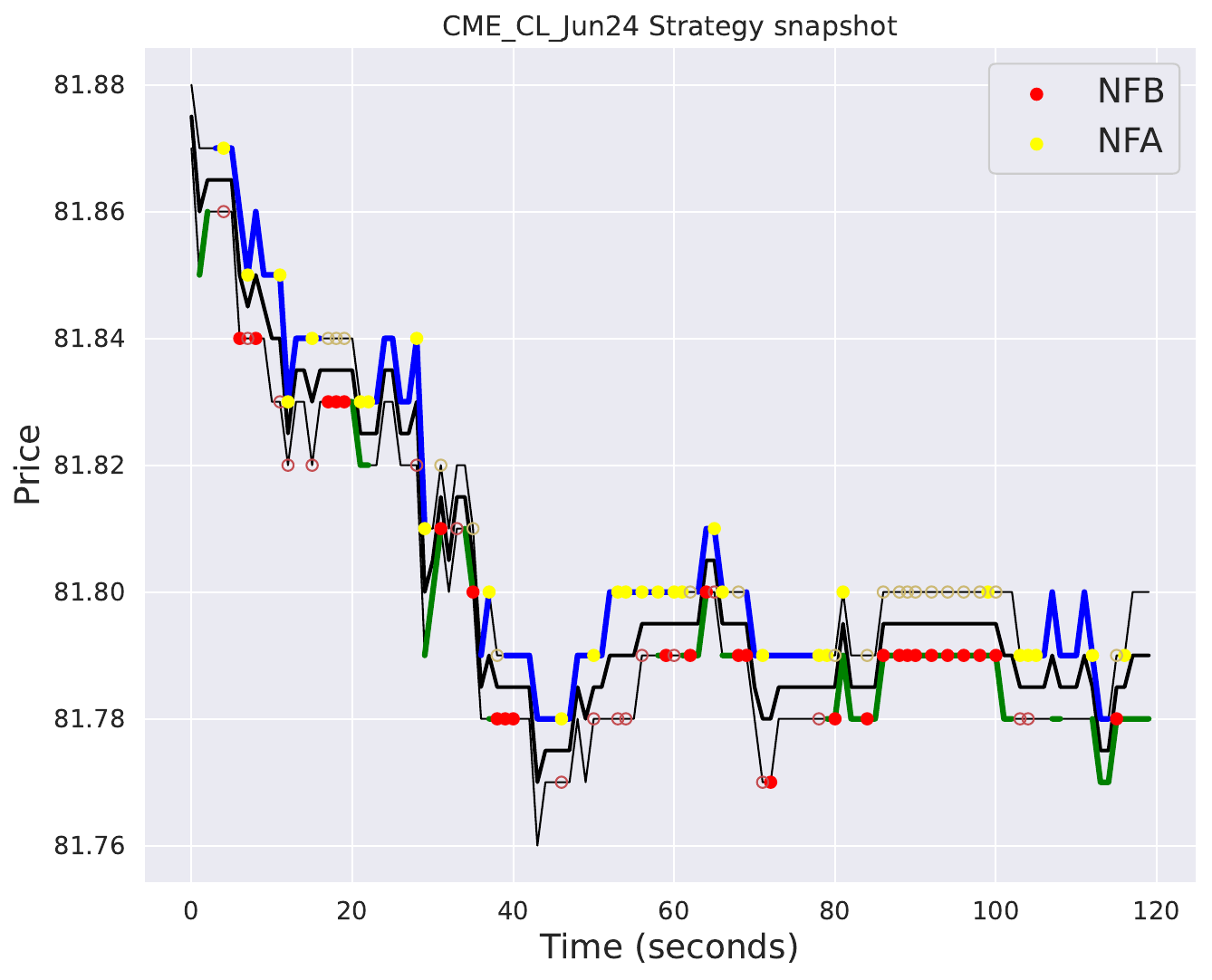} 
\end{subfigure}
\begin{subfigure}{0.5\textwidth}
\includegraphics[width=0.9\linewidth, height=6cm]{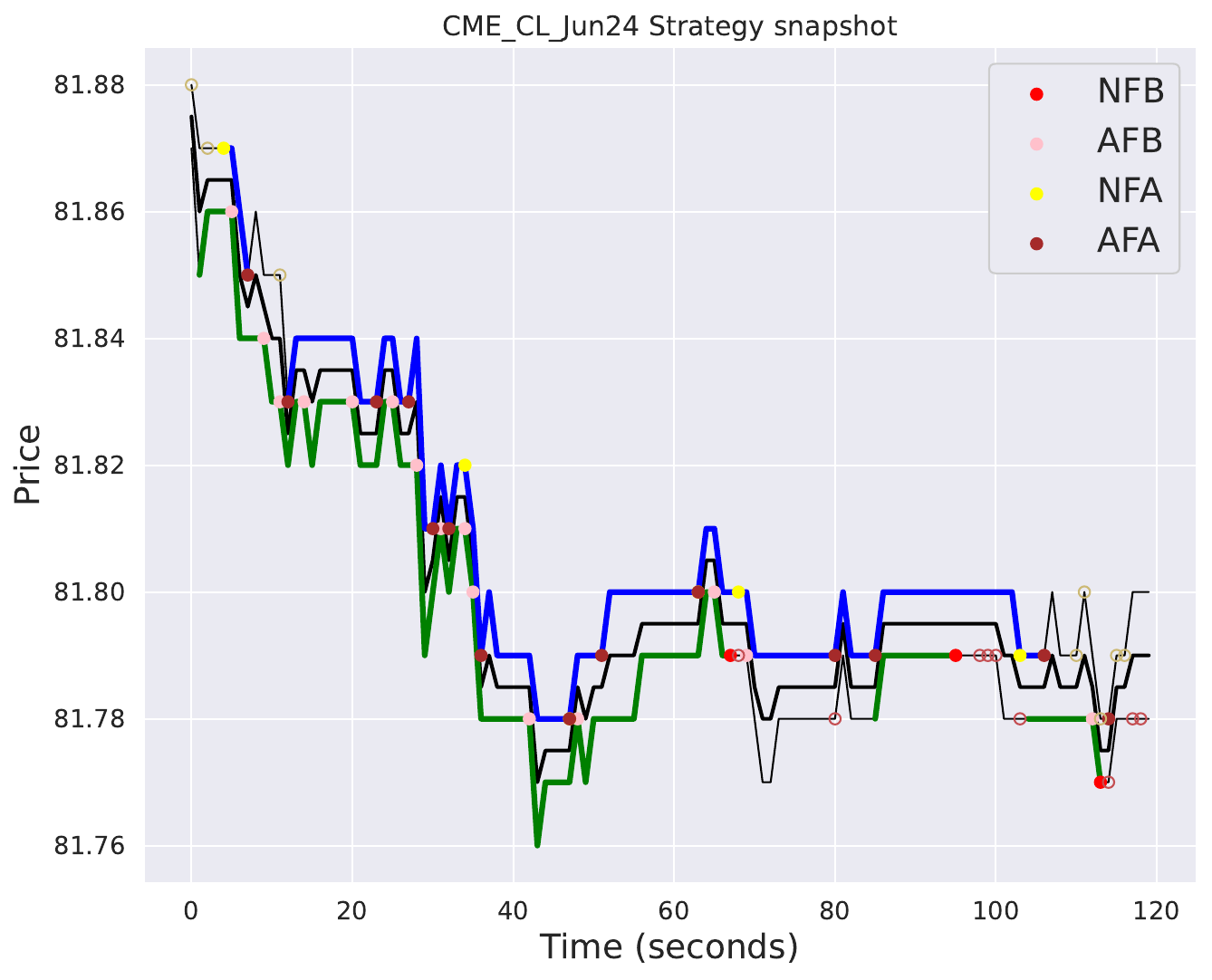}
\end{subfigure}
\captionsetup{font=small}
\caption{A snapshot of a random strategy path in the benchmark (left) and improved (right) simulation environments. The top (black/blue), middle (black), and bottom (black/green) lines represent the best ask, midprice, and best bid, respectively. Blue (green) segments indicate the agent is posted on the best ask (bid). Closed circles show fills received, while open circles represent fills the MM would have received if posted.}
\label{fig:figure3}
\end{figure}

Now, we will review the strategy simulation from our improved simulation environment alongside the performance of the benchmark environment. In Figure \ref{fig:figure3}, we begin by showing a snapshot of the strategy over a random 120 second path in CL, where one can see when the MM is posted on the best bid/ask and when they receive trade order fills. The left and right figures show the strategy simulation in the benchmark and improved environments, respectively.   Here, the green (blue) lines indicate when the MM is posted on the best bid (ask), the filled circles indicate when the MMs LO is filled on the best bid/ask, and the unfilled circles indicate times the MM would have been filled if they had a LO posted. One noticeable feature, in the figure on the left, is that whenever the agent is posted and the price moves through their order, they do not automatically receive a fill. This, as we mentioned earlier, is because MOs are simulated independently from the price process in the benchmark models in \citet{cartea2015algorithmic} and \citet{jaimungalgit}, and is contrary to what would happen in reality. And so, all the fills in this left figure would be referred to as non-adverse fills. In the right figure, one can see where all the adverse fills would have occurred (denoted by AFB and AFA). Note that this can also alter the posting strategy later on because the inventory process now evolves differently to how it would in the benchmark environment. 

\begin{table}[H]
	\begin{center}
		\begin{tabular}{ |p{3cm}|p{3cm}|p{3cm}|p{3cm}|  }
 		\hline
 		\multicolumn{4}{|c|}{\textbf{Parameters}} \\
 		\hline
 		\textbf{Parameter} & \textbf{Value} 		&\textbf{Parameter} & \textbf{Value}\\
 		\hline
 		$\sigma$ & 0.005 & T & 120 seconds\\
      	POV & 0.2 & Nq & 7 \\
      	$\zeta$ & 0.05 & $\epsilon$ & 0.002 \\
      	$\eta$ & 0.001 & $\Delta$ & 0.01 \\
      	$\varphi$ & 0.01 & $\phi$ & 0\\
      	$\lambda^+$ & 0.5833 & $\lambda^-$ & 0.5833 \\
      	Ndt & 120 & dt & 1 second \\
       
 		\hline
		\end{tabular}
	\captionsetup{font=small}
	\caption{Simulation parameters.}
	\label{tab:MMparams}
	\end{center}
\end{table}

\begin{table}[H]
  \begin{center}
    
    \begin{tabular}{l|c|r} 
      \textbf{Fill Type} & \textbf{Amount}\\
      \hline
      AFA & 3950\\
      NFA & 825\\
      AFB & 3943\\
      NFB & 813
    \end{tabular}
    \captionsetup{font=small}
    \caption{Number of trade order fills that were adverse and non-adverse over 330 simulations throughout the trading on April 24th, 2025. }
    \label{tab:fillssim}
  \end{center}
\end{table}

It is quite apparent that the timing of the trade order fills has changed and many trade order fills occur at very unfavorable prices. Earlier, in our simple trading example, where we gave results in Table \ref{tab:tablemotivex}, we saw there that a overwhelming majority of the fills for the strategy simulation in CL were adverse, as well as for ES, NQ and ZN. This test was also conducted on the exact same trading day as this simulation. In Table \ref{tab:fillssim}, we show the number of each type of fill that occurred within our improved simulation environment. We can see that using a much lower non-adverse fill probability of $\rho=0.2$ created a similar distribution of fills between non-adverse fills and adverse fills as in Table \ref{tab:tablemotivex}, and we will soon see that the overall performance has now dropped significantly. This trend also persists in the strategy simulations using ES, NQ and ZN data, as can be seen in the appendix. Based on our empirical evidence, we believe that these results better depict how the strategy would perform in reality. However, we may not want to be overly pessimistic by thinking the SOC MM strategy should have the same number of adverse fills as the basic posting strategy, because the SOC MM model has a short-term alpha predictor, which could potentially predict some of these adverse fills. However, it is highly unlikely that any short-term alpha predictor could anticipate all future adverse fills. The simple predictor in Equation \eqref{eq: dalpcart} lacks the sophistication needed for such predictions, especially given its simplicity and the likely alpha decay it has undergone due to its public availability. Recall our earlier note on the alpha decay of a publicly disclosed trading strategy in Section 3.2. 

\begin{figure}[H]
        \centering
        \begin{subfigure}[b]{0.4\textwidth}
            \centering
            \includegraphics[width=\textwidth]{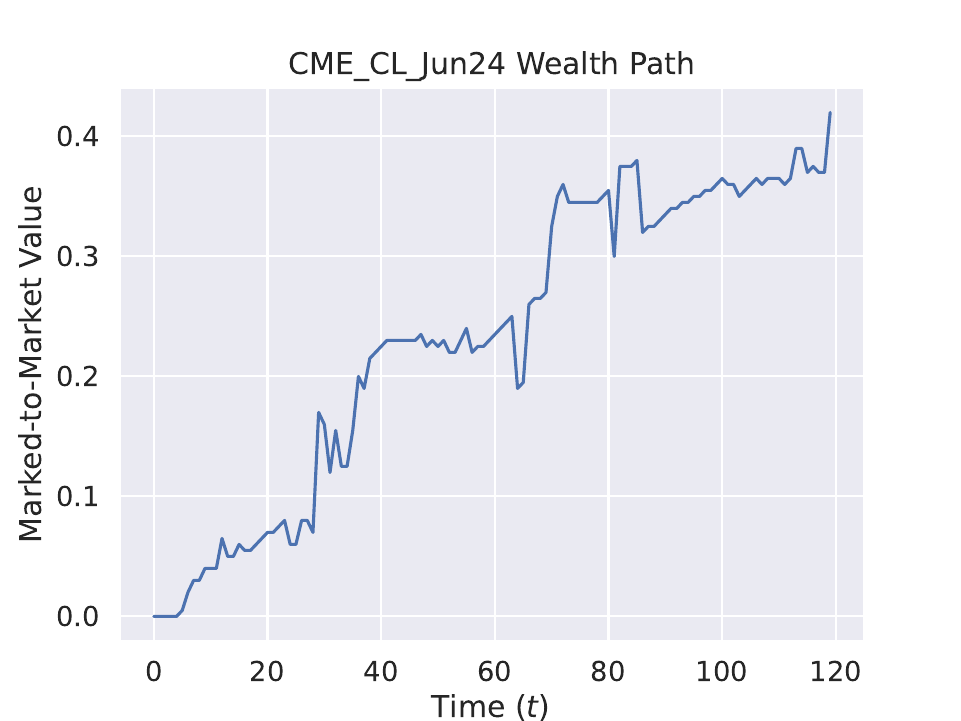}
        \end{subfigure}
        \hfill
        \begin{subfigure}[b]{0.4\textwidth}  
            \centering 
            \includegraphics[width=\textwidth]{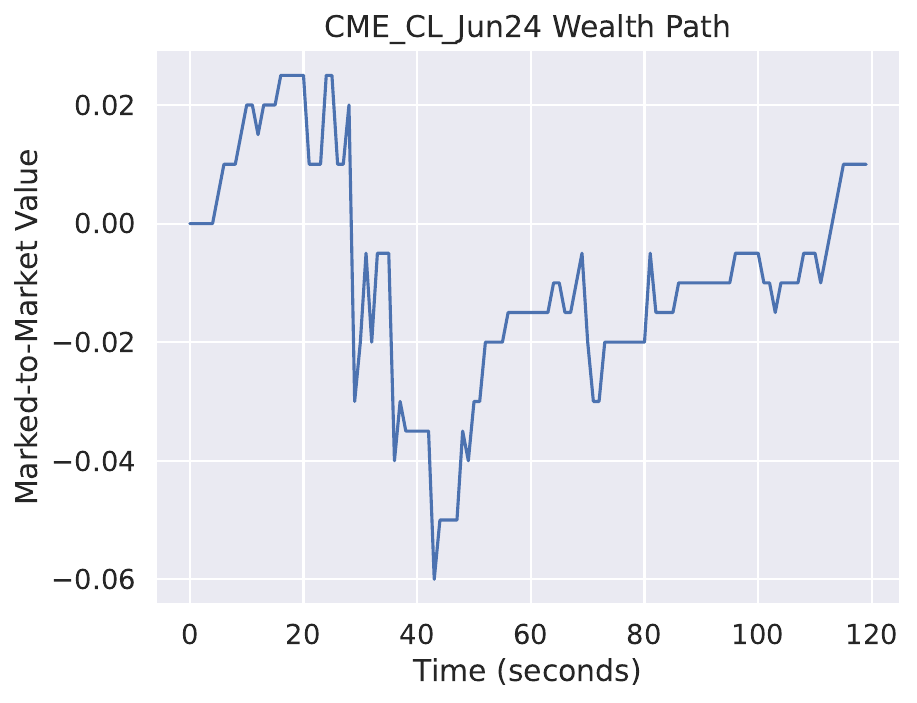}
        \end{subfigure}
        \vskip\baselineskip
        \begin{subfigure}[b]{0.4\textwidth}   
            \centering 
            \includegraphics[width=\textwidth]{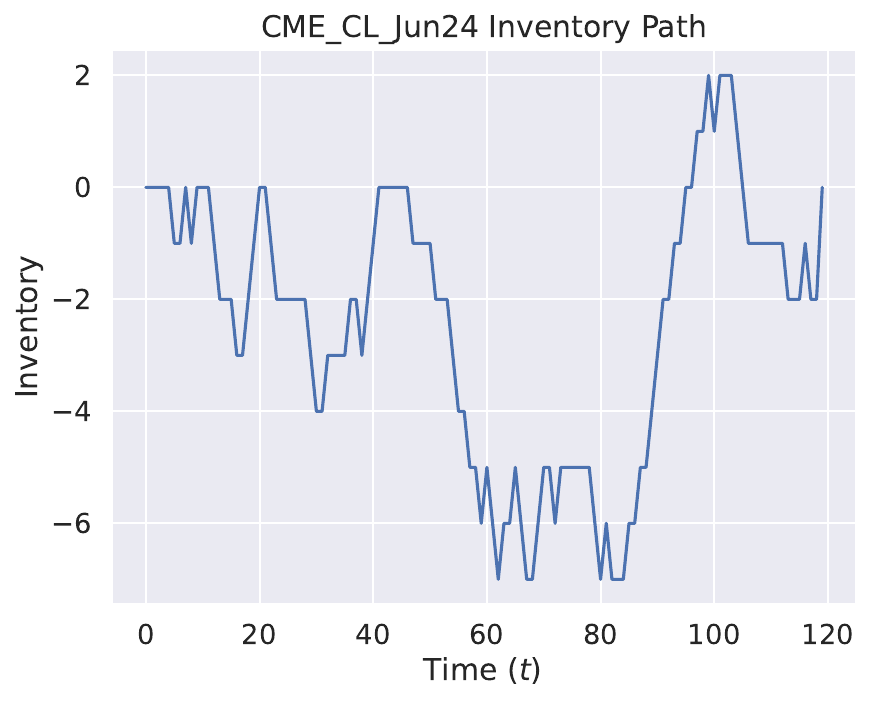}
        \end{subfigure}
        \hfill
        \begin{subfigure}[b]{0.4\textwidth}   
            \centering 
            \includegraphics[width=\textwidth]{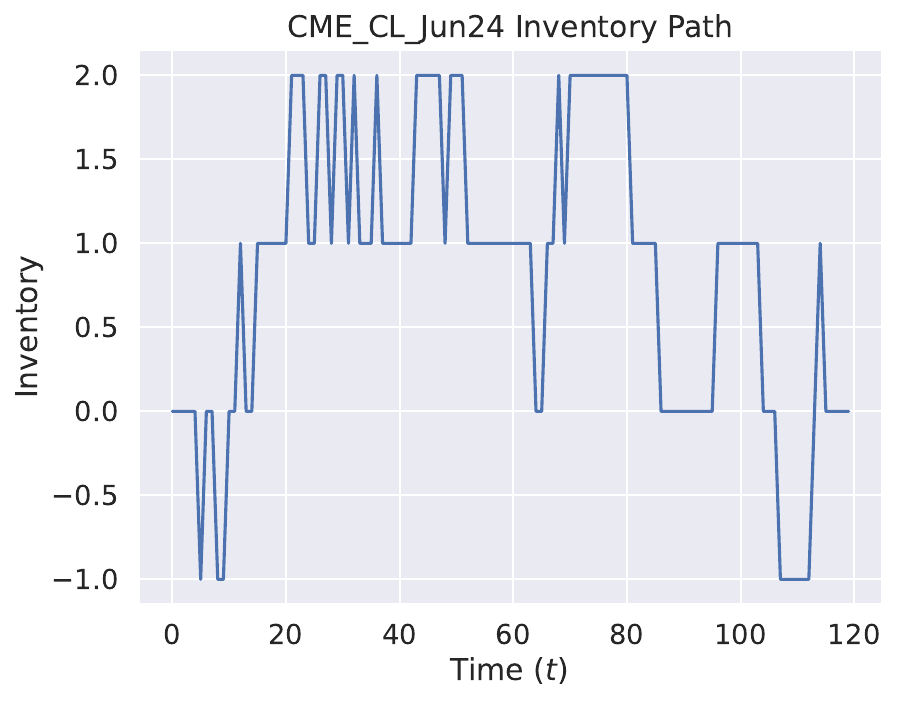}
        \end{subfigure}
        \captionsetup{font=small}
        \caption{ Wealth and inventory processes for the same random path of the strategy under the benchmark (left) and improved (right) simulation environments.}
        \label{fig: wealthinvsim}
    \end{figure}

Next, in Figure \ref{fig: wealthinvsim}, one can see the wealth and inventory path for the same random 120-second run of the strategy. Here, again, the left and right figures are in the benchmark and improved simulation environments, respectively. The wealth path is marked-to-market at each time step as in Equation \eqref{eq: cashcartea} and the inventory path changes whenever a trade order fill is received as in Equation \eqref{eq: invcartea}. Here, the MM's wealth path in the left figure ends positive and at a higher value than in the right figure, both over the same 120 second price path. The inventory paths show that the benchmark environment traded quite differently from our improved environment, which is largely due to the additional adverse fills this trader received, but also the lower non-adverse fill probability. 
Thus, it is quite apparent that these paths now evolve very differently. Lower non-adverse fill probabilities and adverse fills have significantly affected how these processes evolved. 

\begin{figure}[H]

\begin{subfigure}{0.5\textwidth}
\includegraphics[width=0.9\linewidth, height=6cm]{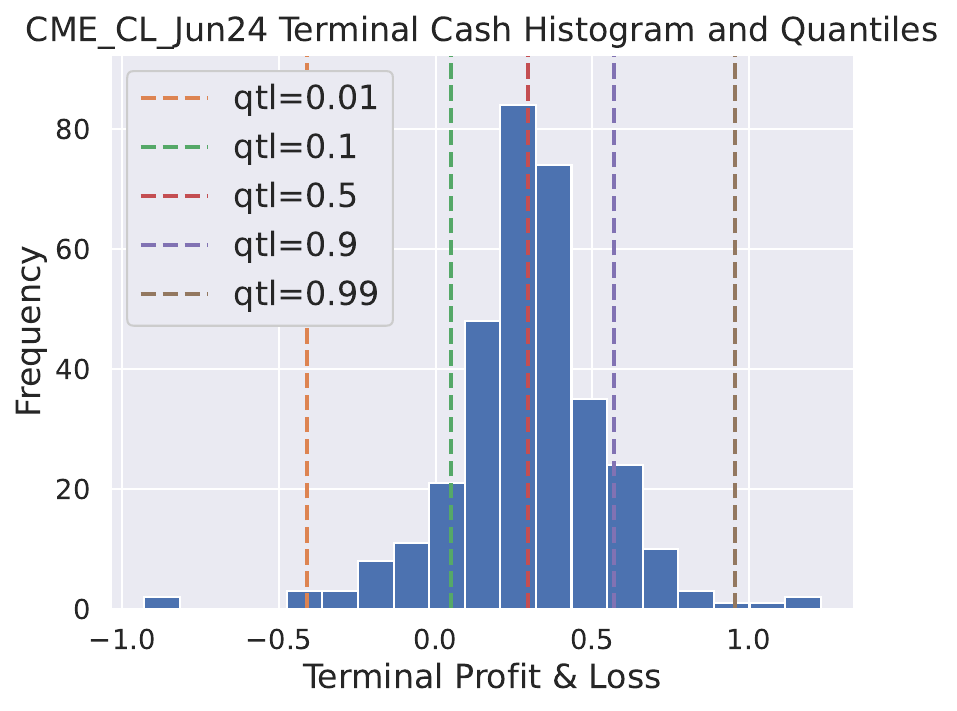} 
\end{subfigure}
\begin{subfigure}{0.5\textwidth}
\includegraphics[width=0.9\linewidth, height=6cm]{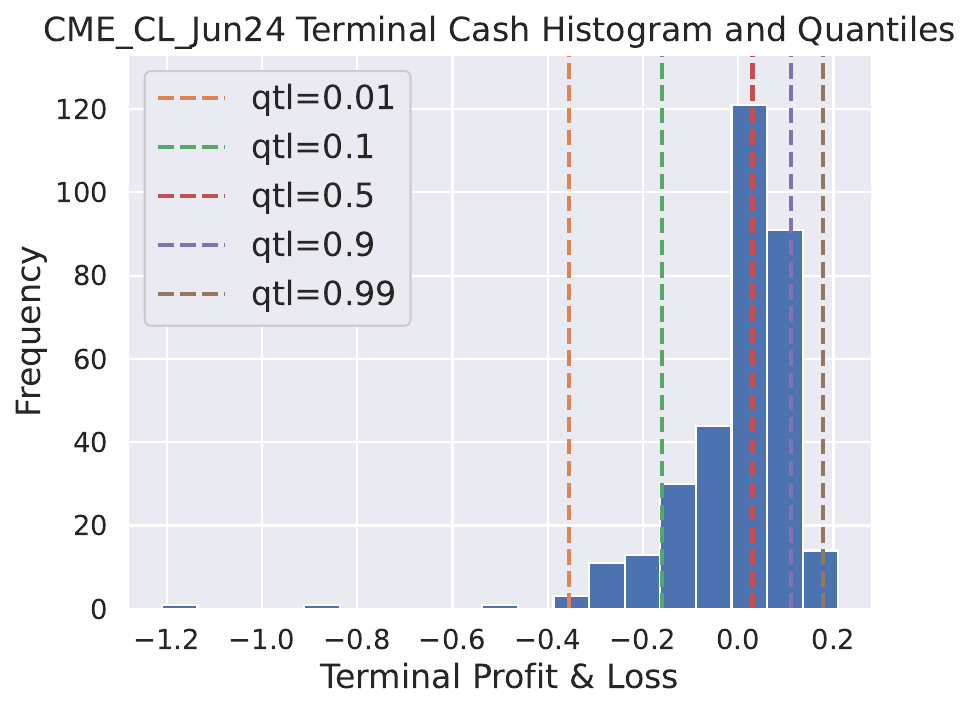}
\end{subfigure}
\captionsetup{font=small}
\caption{Terminal cash histogram for the benchmark (left) and improved (right) simulation environments over all 330 simulation paths.}
\label{fig:HistSim}
\end{figure}

In order to tell the full story, we show a histogram, in Figure \ref{fig:HistSim}, of the terminal cash values from all 330 simulation runs of the strategy over this trading day. Each bin indicates how often the strategy paths attained a certain P\&L. Here, again, the left and right figures show the benchmark and improved simulation environments, respectively. One can see that, overall, the strategy performed reasonably well in the benchmark environment and much worse in our improved environment. We believe the performance metric in the benchmark environment is over-inflated, as the MM received a lot more non-adverse fills than likely in reality, as well not receiving any adverse fills, which clearly significantly affects the performance.  
This confirms to us that simulating the performance of a short-term style trading strategy, in particular one that involves posting many LOs, is very misleading if it does not track adverse fills and include more accurate non-adverse fill probabilities. This is just as evident in ES, NQ and ZN, where similar results for these assets can be seen in the appendix.

\section{Conclusion and Future Recommendations}
In this paper, we simulated the performance of an SOC optimal MM strategy under different fill assumptions, including the presence or absence of adverse fills and varying probabilities of non-adverse fills. Since many MM strategies rely heavily on posting LOs throughout the LOB, our findings are particularly relevant to these strategies, as well as to any trading strategy that posts large numbers of LOs. While simplifying assumptions are often necessary in mathematical finance models, some can materially distort performance evaluation. In our setting, modeling price dynamics and MO arrivals independently leads to significant inaccuracies. In particular, excluding adverse fills and realistic non-adverse fill probabilities substantially inflates trading performance.

Adverse fills are unavoidable in LO posting strategies, especially in short-term trading environments. Although short-term alpha predictors such as those proposed by \citet{cartea2015algorithmic} and \citet{cartea2018algorithmic} may help anticipate adverse moves, large orders can still enter unexpectedly and rapidly remove price levels before orders can be cancelled, even for HFTs with speed advantages.

Future research could extend the model by introducing a more dynamic MM framework in which key parameters evolve over time. For example, MO sizes could vary instead of being fixed at one unit, which may better capture the clustering effects commonly observed in LOB data and improve the modeling of adverse fills and non-adverse fill probabilities. Another extension would be to adapt the posting strategy to asset-specific dynamics, such as spreading LO placements further apart for highly volatile assets like NQ. Recent work by \citet{arroyo2024deep} and \citet{maglaras2022deep} suggests that deep learning methods may provide promising approaches for estimating fill probabilities in more dynamic settings. The performance of short-term alpha predictors should also be examined more carefully, since they are a central component of the framework but are unlikely to achieve such simple predictive accuracy in practice.

\section*{Acknowledgements}
We would like to thank MITACS and NSERC for research funding. We would specifically like to thank Futures First for collaborating in this MITACS project, where we would also like to give a special mention to Timothy DeLise (PhD candidate at the University of Montreal and former MITACS participant in projects at Futures First) and Myles Sjogren (Quantitative Analyst at Futures First) in helping to set-up parts of the coding framework for deploying the basic posting algorithm described in Section 3.2.

\section*{Declarations of Interest}
The authors declare no conflicts of interest. 

\bibliographystyle{apalike}
\bibliography{Paper_1}

\appendix
\section{Appendix: ES, NQ and ZN Strategy Simulations}


\begin{figure}[H]
        \centering
        \begin{subfigure}[b]{0.4\textwidth}
            \centering
            \includegraphics[width=\textwidth]{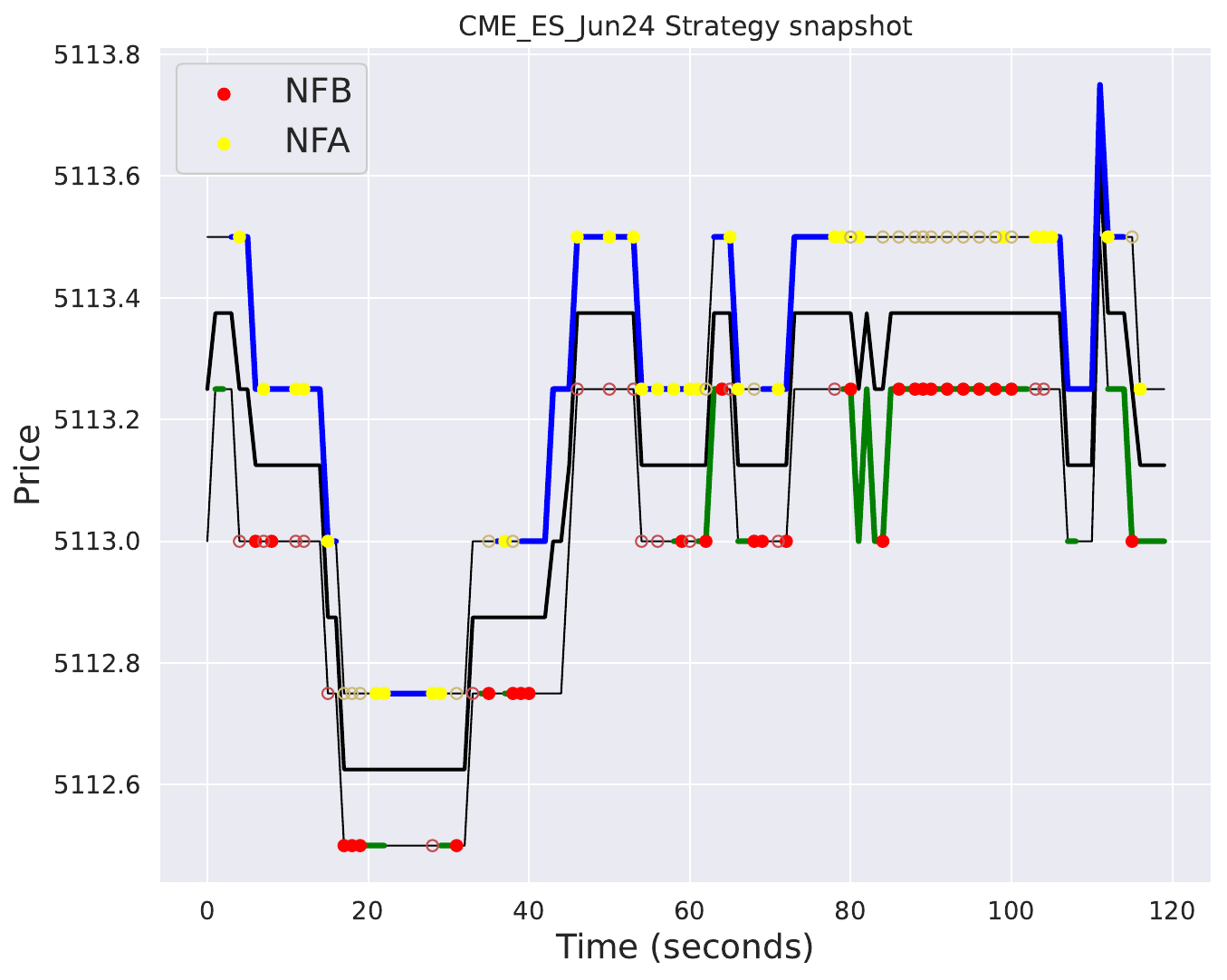}
        \end{subfigure}
        \hfill
        \begin{subfigure}[b]{0.4\textwidth}  
            \centering 
            \includegraphics[width=\textwidth]{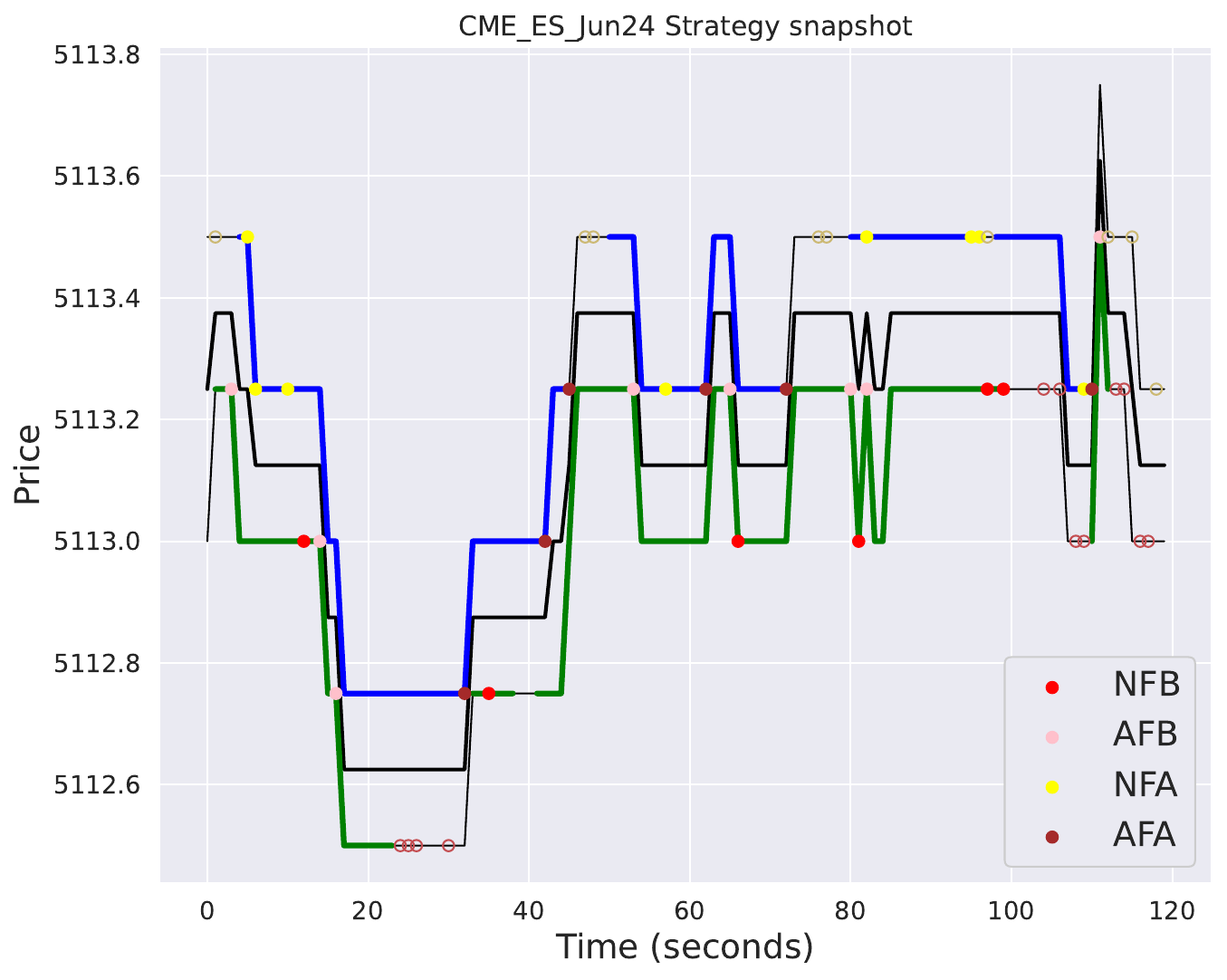}
        \end{subfigure}
        \vskip\baselineskip
        \begin{subfigure}[b]{0.4\textwidth}   
            \centering 
            \includegraphics[width=\textwidth]{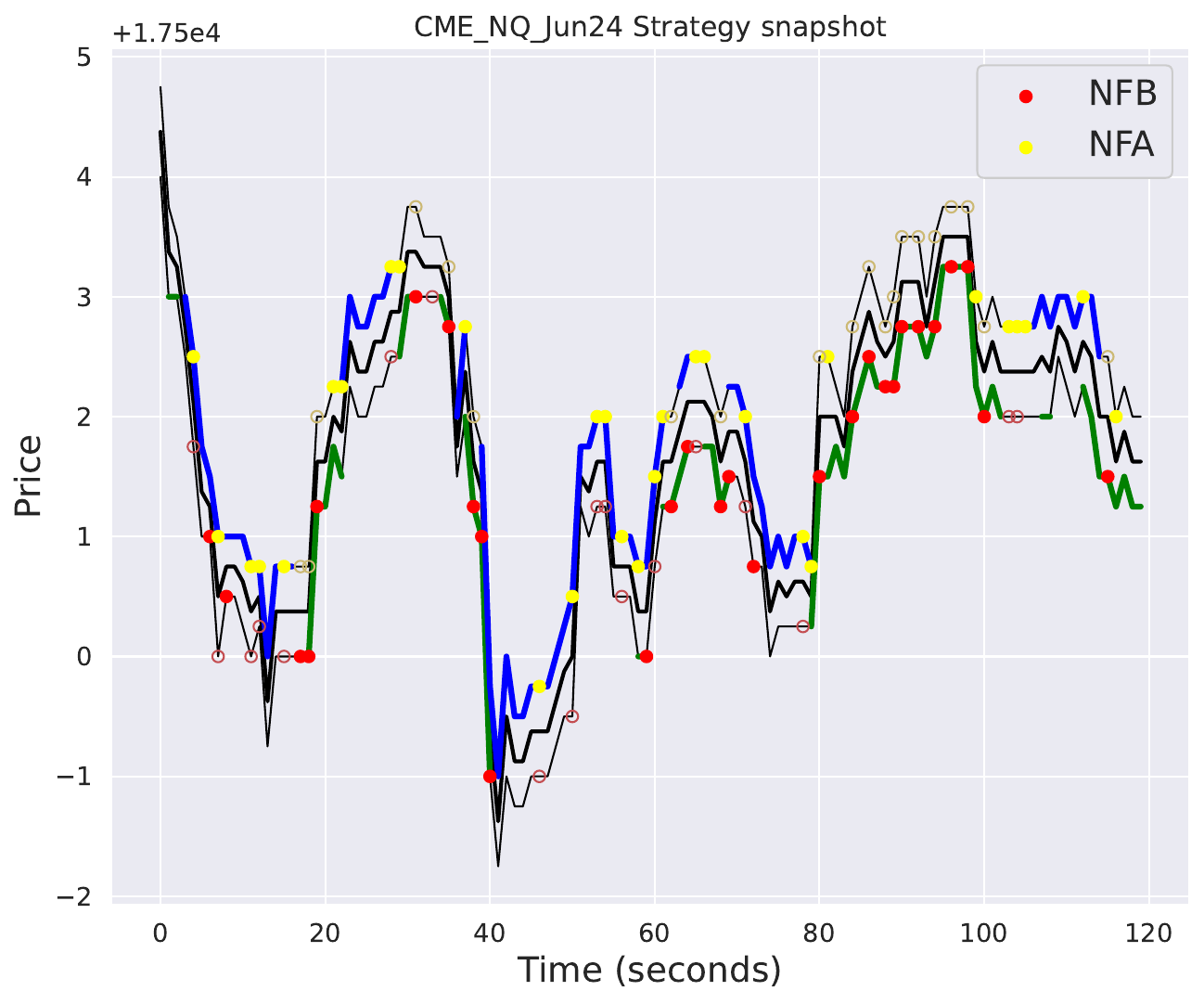}
        \end{subfigure}
        \hfill
        \begin{subfigure}[b]{0.4\textwidth}   
            \centering 
            \includegraphics[width=\textwidth]{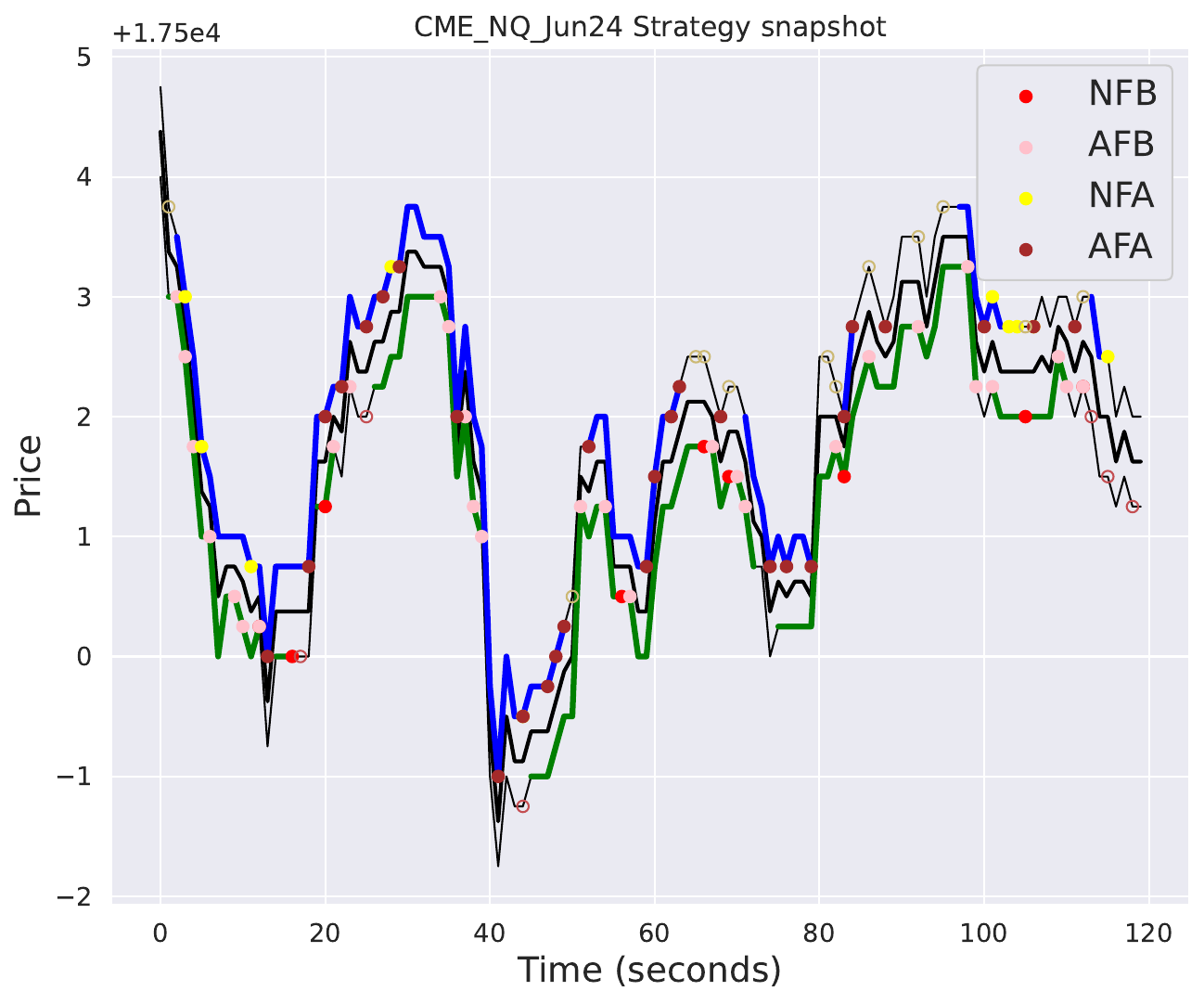}
        \end{subfigure}
        \vskip\baselineskip
        \begin{subfigure}[b]{0.4\textwidth}   
            \centering 
            \includegraphics[width=\textwidth]{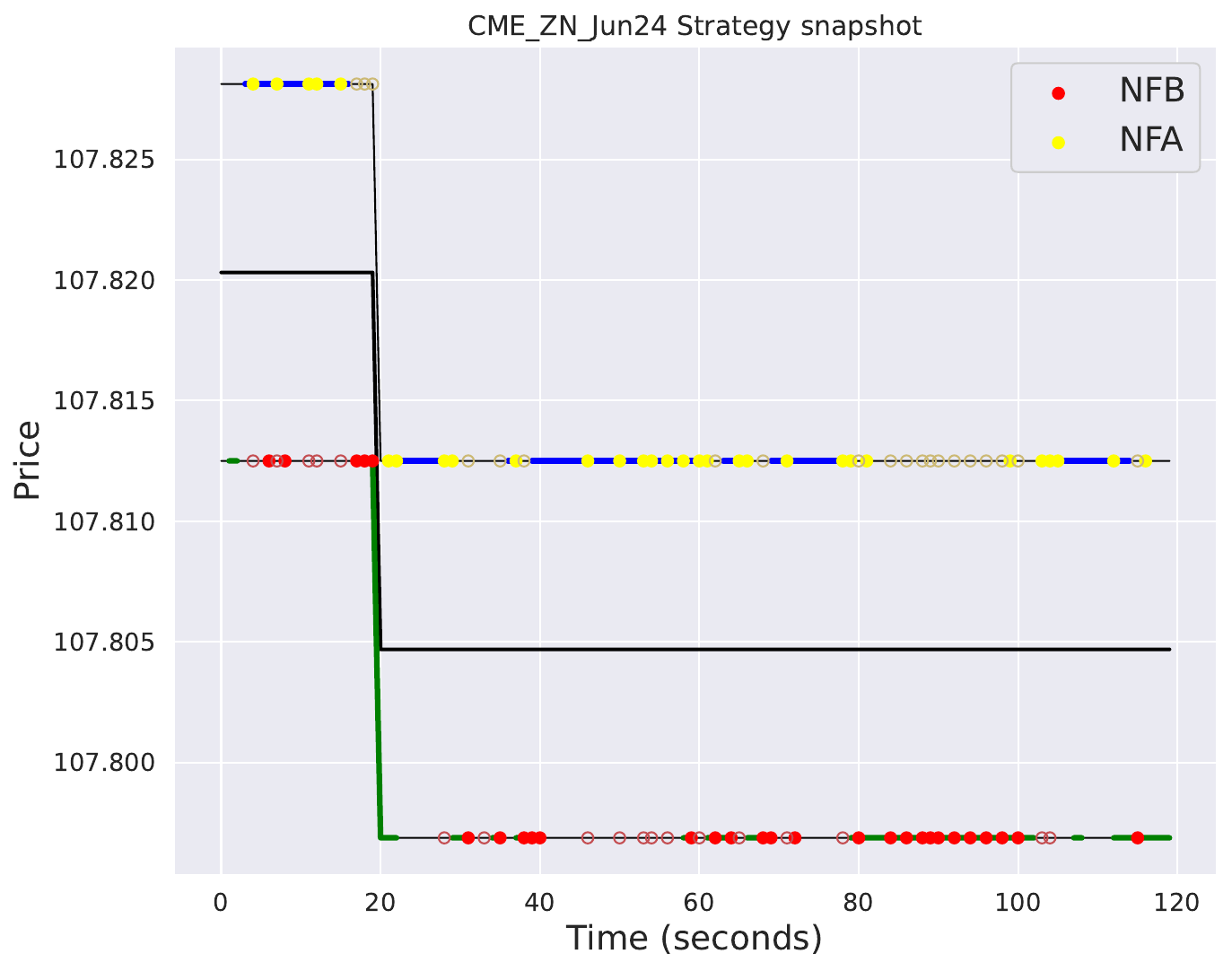}
        \end{subfigure}
        \hfill
        \begin{subfigure}[b]{0.4\textwidth}   
            \centering 
            \includegraphics[width=\textwidth]{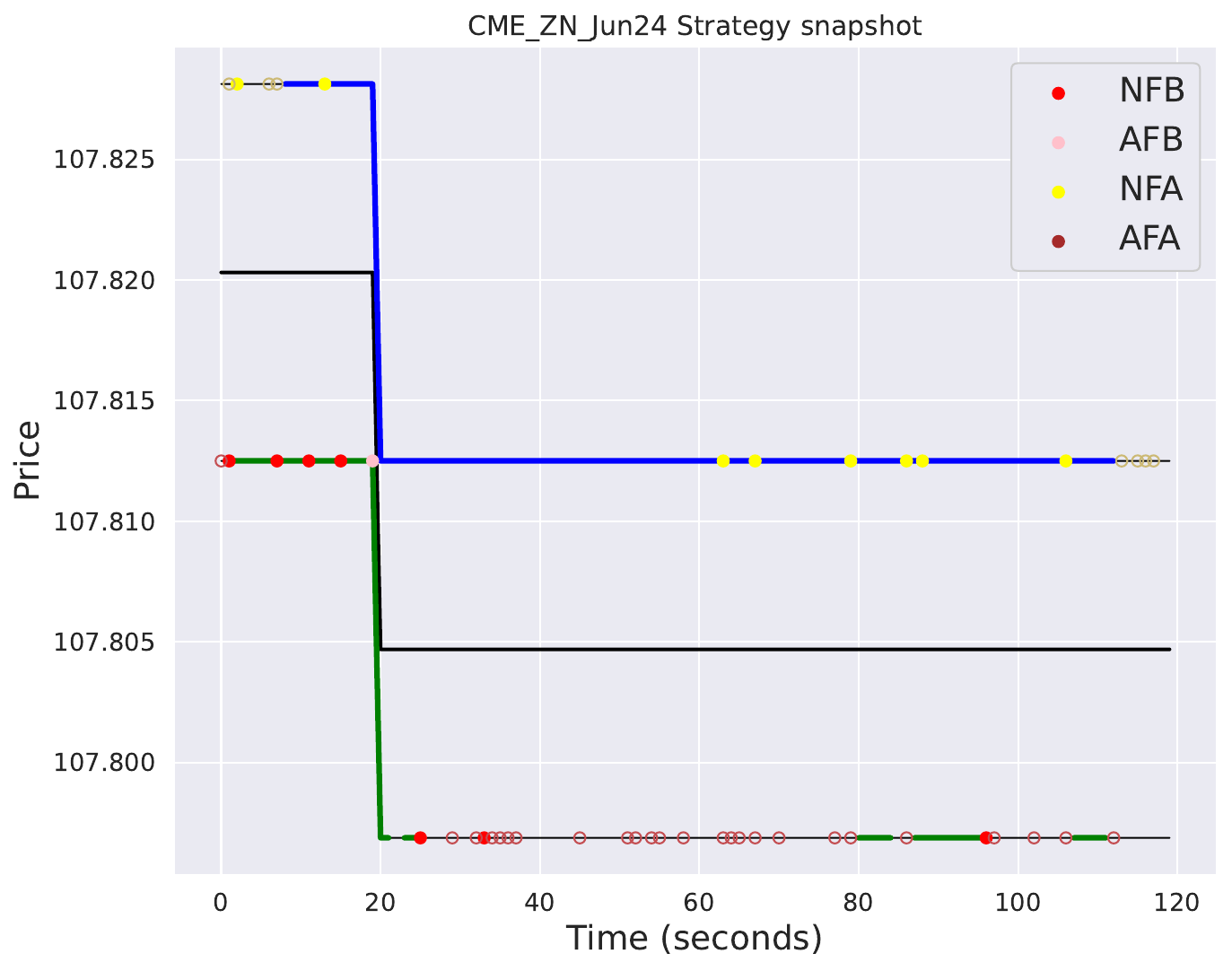}
        \end{subfigure}
        \captionsetup{font=small}
        \caption{ A snapshot of the same random path of the strategy in the benchmark (left) and improved (right) simulation environments. The top line (black/blue) indicates the best ask, the middle line (black) the midprice and the bottom line (black/green) the best bid. When the top (bottom) line is blue (green), the agent is posted on the best ask (bid).  Closed circles indicate fills received when the market-maker is posted and open circles indicate fills that the market-maker would've received if they were posted. }
    \end{figure}

\begin{figure}[H]
        \centering
        \begin{subfigure}[b]{0.4\textwidth}
            \centering
            \includegraphics[width=\textwidth]{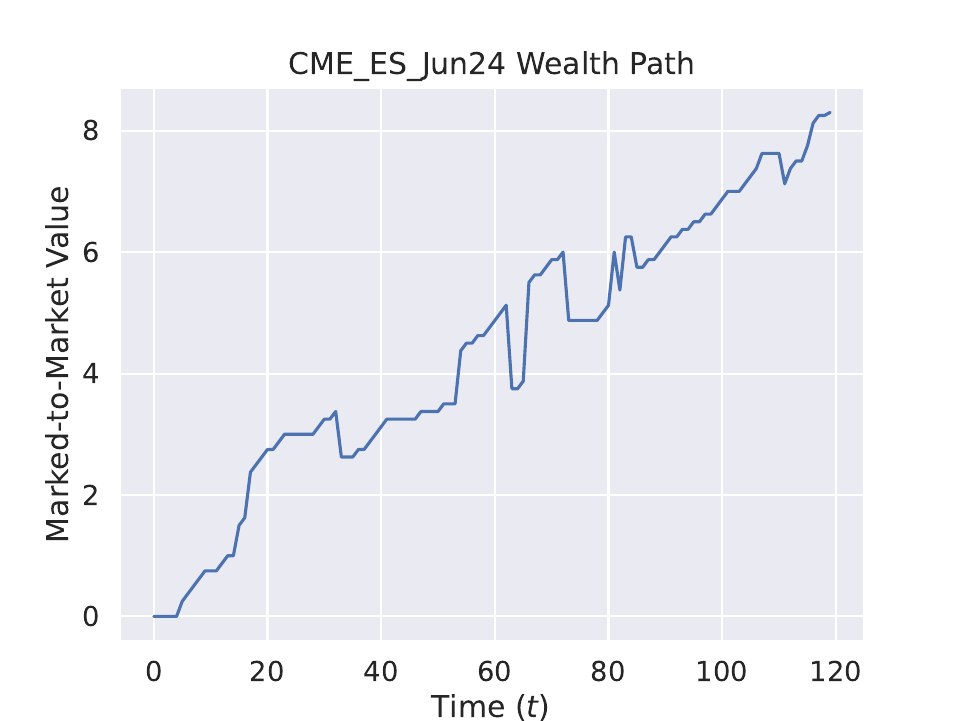}
        \end{subfigure}
        \hfill
        \begin{subfigure}[b]{0.4\textwidth}  
            \centering 
            \includegraphics[width=\textwidth]{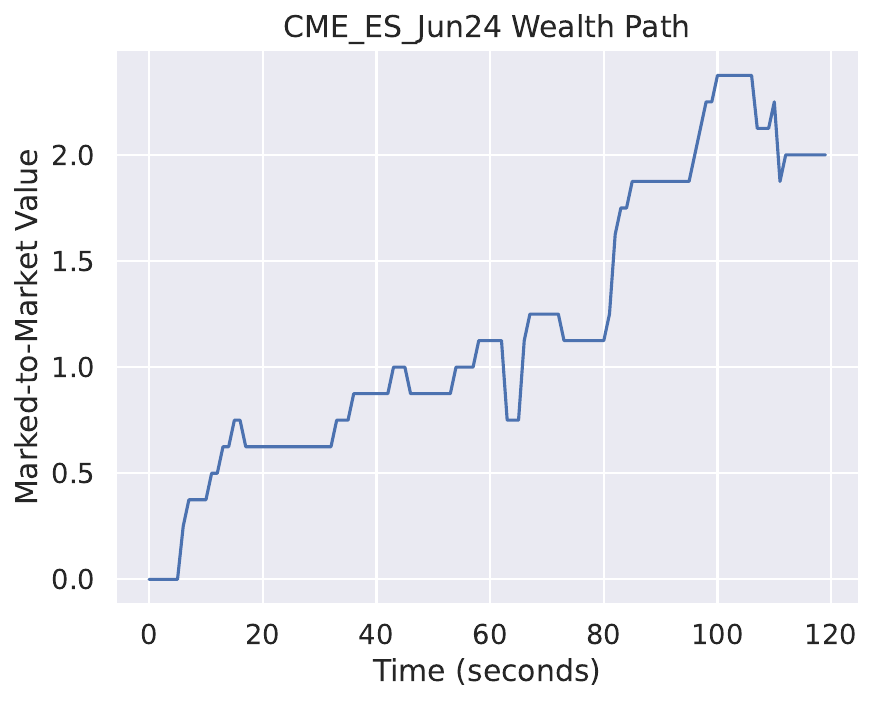}
        \end{subfigure}
        \vskip\baselineskip
        \begin{subfigure}[b]{0.4\textwidth}   
            \centering 
            \includegraphics[width=\textwidth]{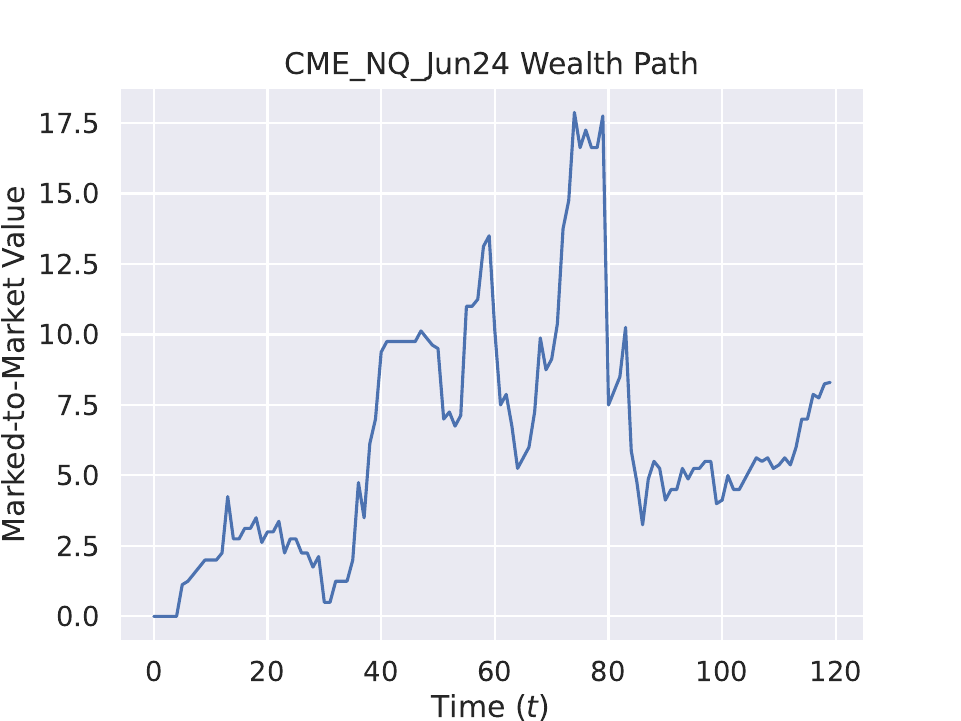}
        \end{subfigure}
        \hfill
        \begin{subfigure}[b]{0.4\textwidth}   
            \centering 
            \includegraphics[width=\textwidth]{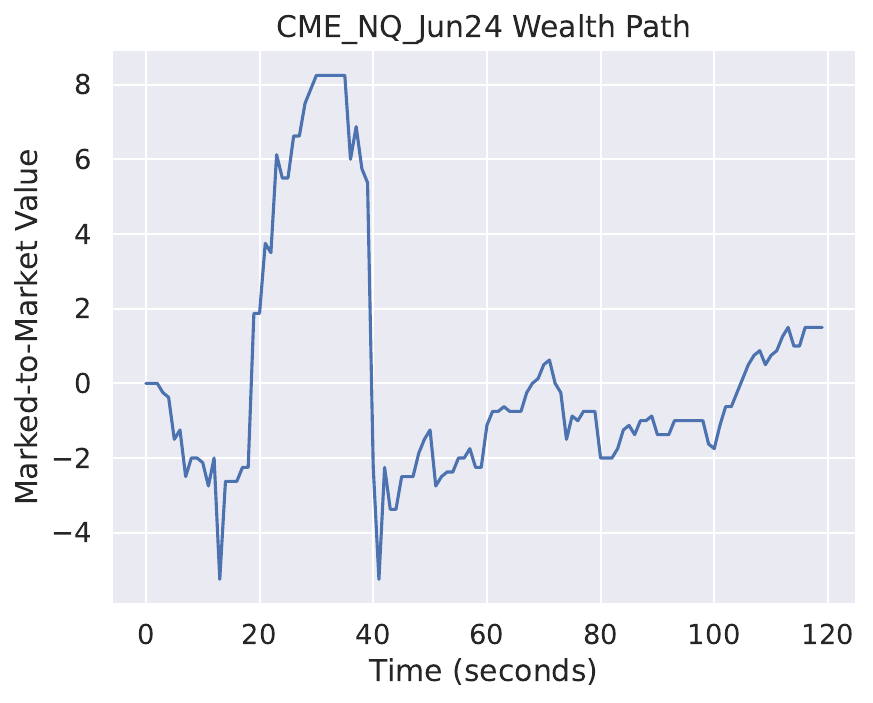}
        \end{subfigure}
        \vskip\baselineskip
        \begin{subfigure}[b]{0.4\textwidth}   
            \centering 
            \includegraphics[width=\textwidth]{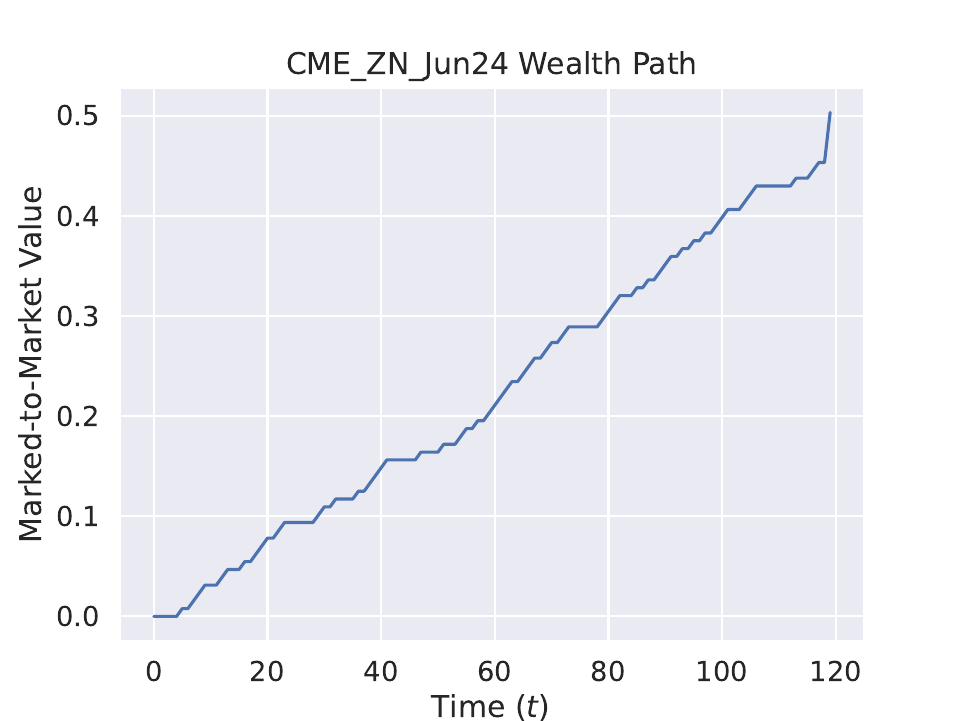}
        \end{subfigure}
        \hfill
        \begin{subfigure}[b]{0.4\textwidth}   
            \centering 
            \includegraphics[width=\textwidth]{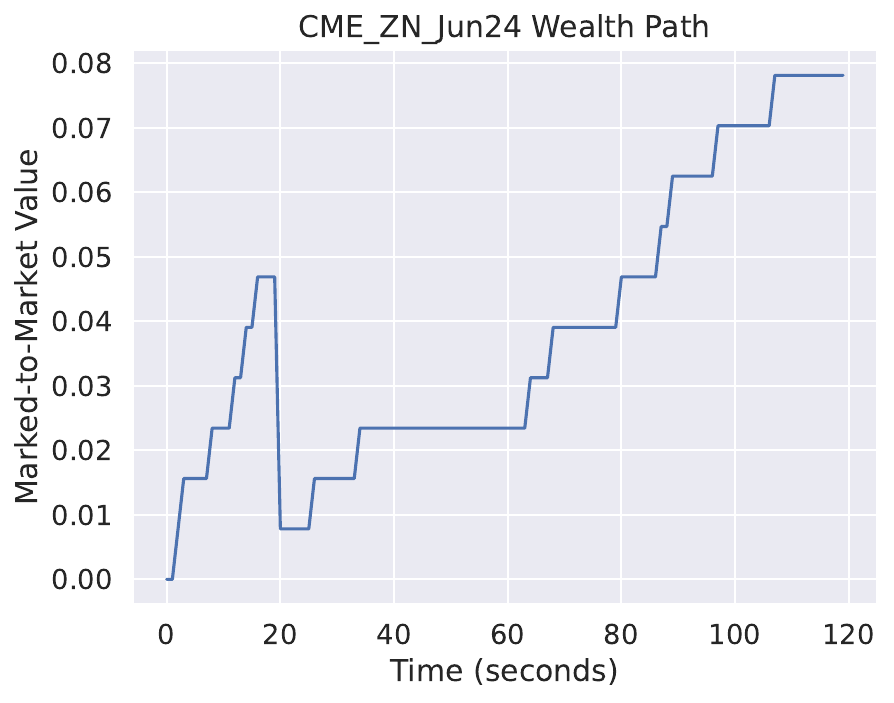}
        \end{subfigure}
        \captionsetup{font=small}
        \caption{Wealth process for the same random path of the strategy under the benchmark (left) and improved (right) simulation environments.}
    \end{figure}

\begin{figure}[H]
        \centering
        \begin{subfigure}[b]{0.4\textwidth}
            \centering
            \includegraphics[width=\textwidth]{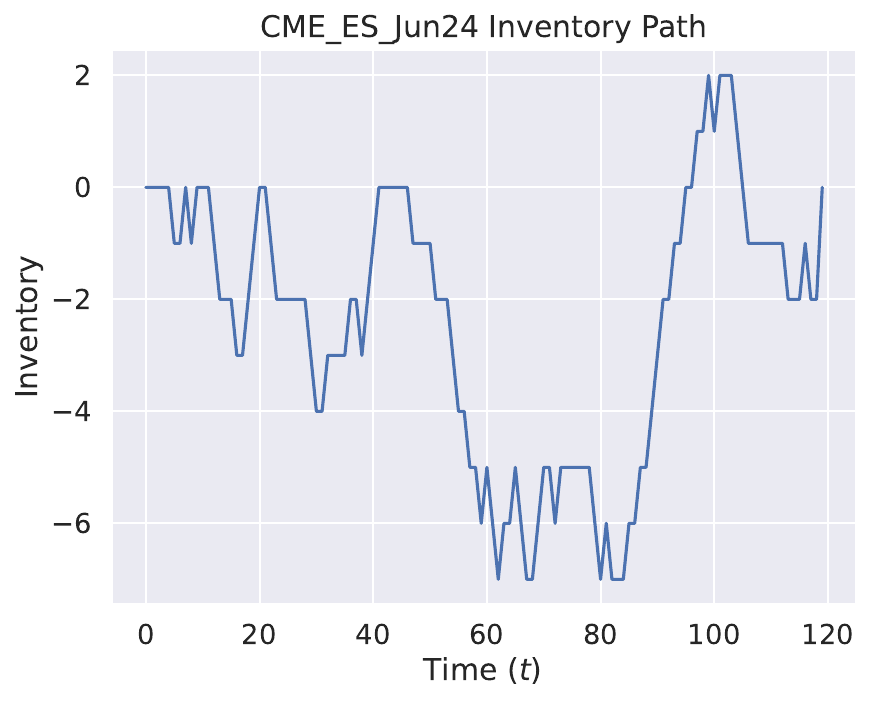}
        \end{subfigure}
        \hfill
        \begin{subfigure}[b]{0.4\textwidth}  
            \centering 
            \includegraphics[width=\textwidth]{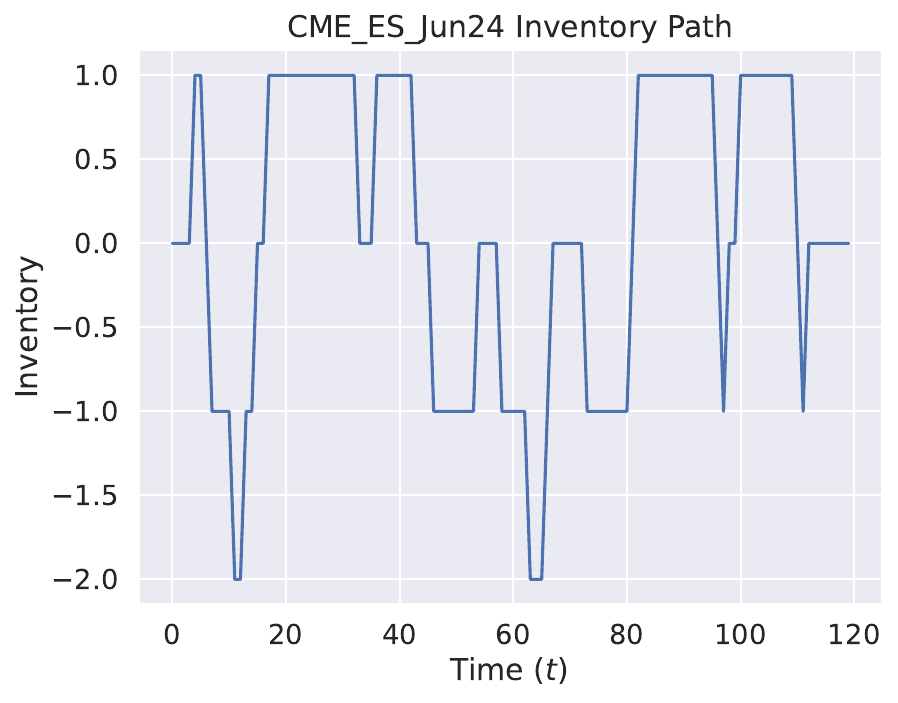}
        \end{subfigure}
        \vskip\baselineskip
        \begin{subfigure}[b]{0.4\textwidth}   
            \centering 
            \includegraphics[width=\textwidth]{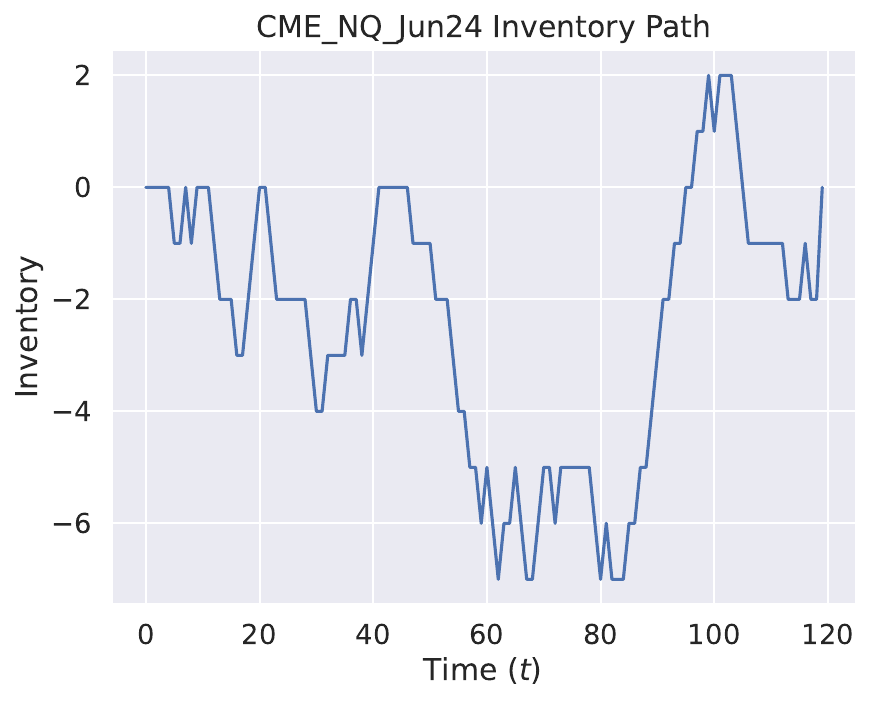}
        \end{subfigure}
        \hfill
        \begin{subfigure}[b]{0.4\textwidth}   
            \centering 
            \includegraphics[width=\textwidth]{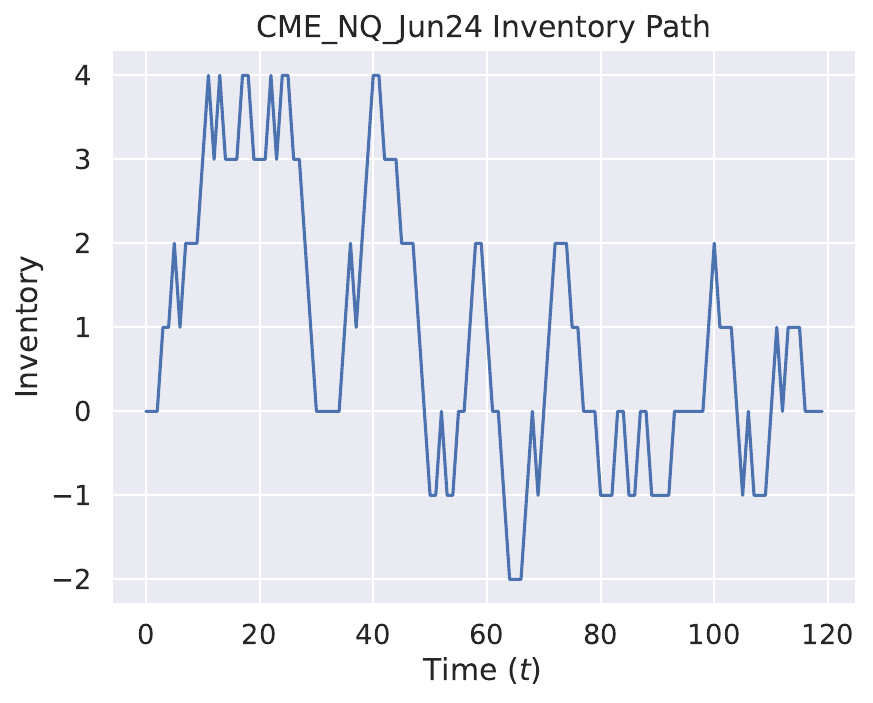}
        \end{subfigure}
        \vskip\baselineskip
        \begin{subfigure}[b]{0.4\textwidth}   
            \centering 
            \includegraphics[width=\textwidth]{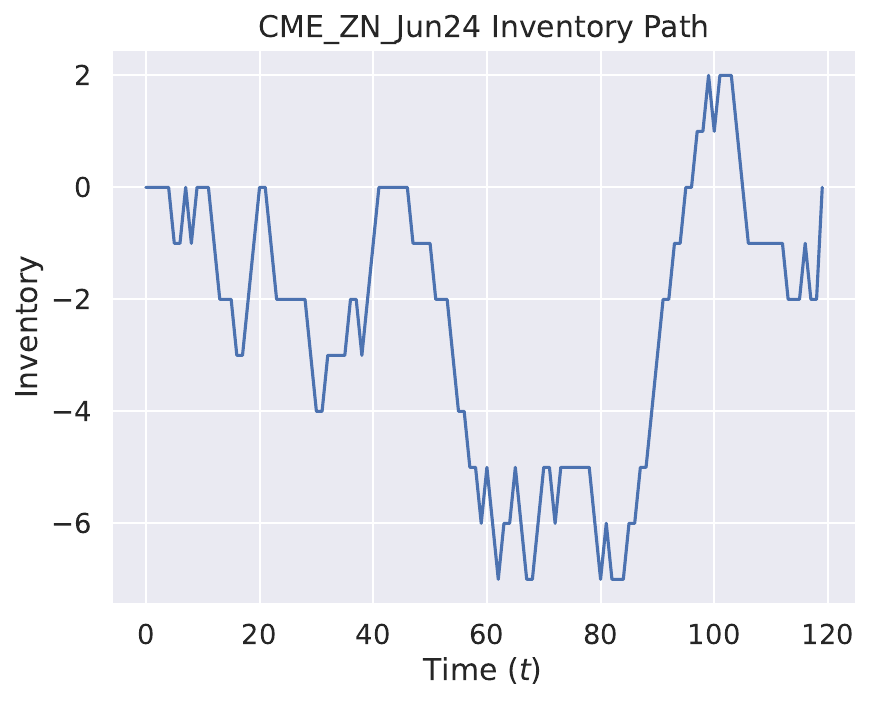}
        \end{subfigure}
        \hfill
        \begin{subfigure}[b]{0.4\textwidth}   
            \centering 
            \includegraphics[width=\textwidth]{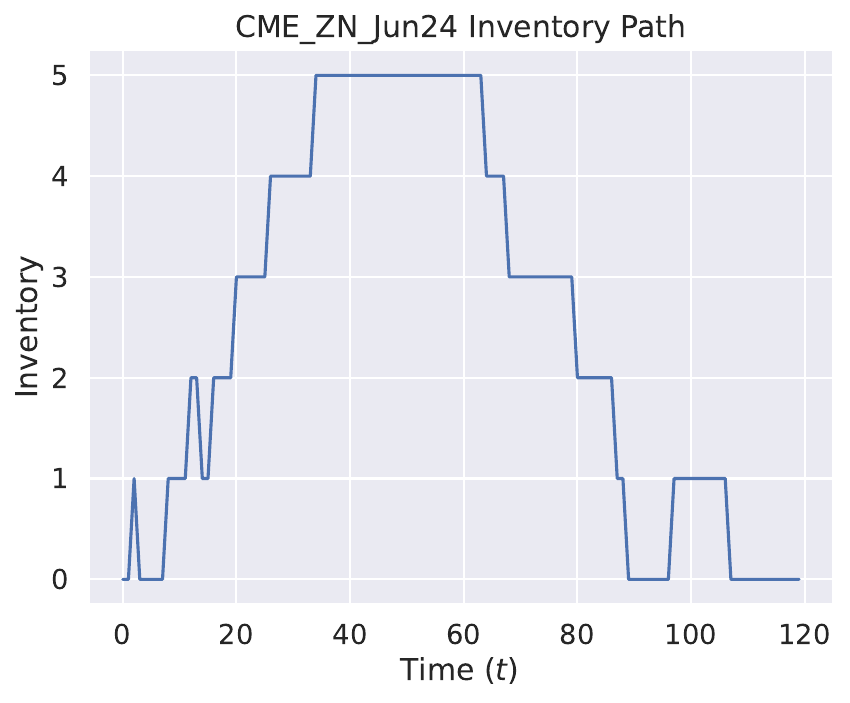}
        \end{subfigure}
        \captionsetup{font=small}
        \caption{ Inventory process for the same random path of the strategy under the benchmark (left) and improved (right) simulation environments.}
    \end{figure}

\begin{figure}[H]
        \centering
        \begin{subfigure}[b]{0.4\textwidth}
            \centering
            \includegraphics[width=\textwidth]{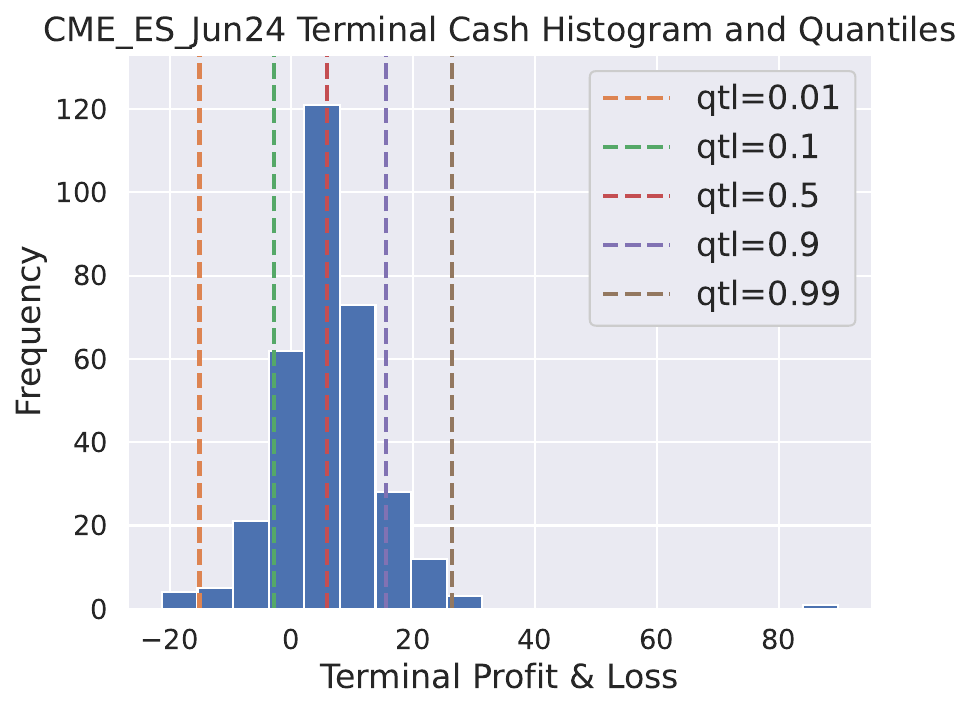}
        \end{subfigure}
        \hfill
        \begin{subfigure}[b]{0.4\textwidth}  
            \centering 
            \includegraphics[width=\textwidth]{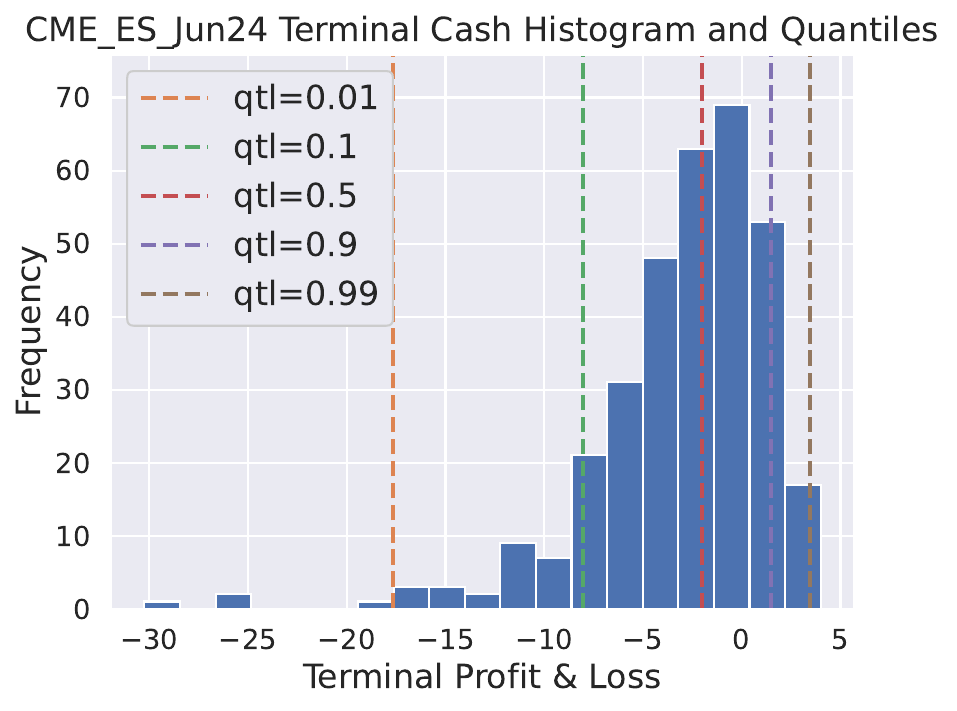}
        \end{subfigure}
        \vskip\baselineskip
        \begin{subfigure}[b]{0.4\textwidth}   
            \centering 
            \includegraphics[width=\textwidth]{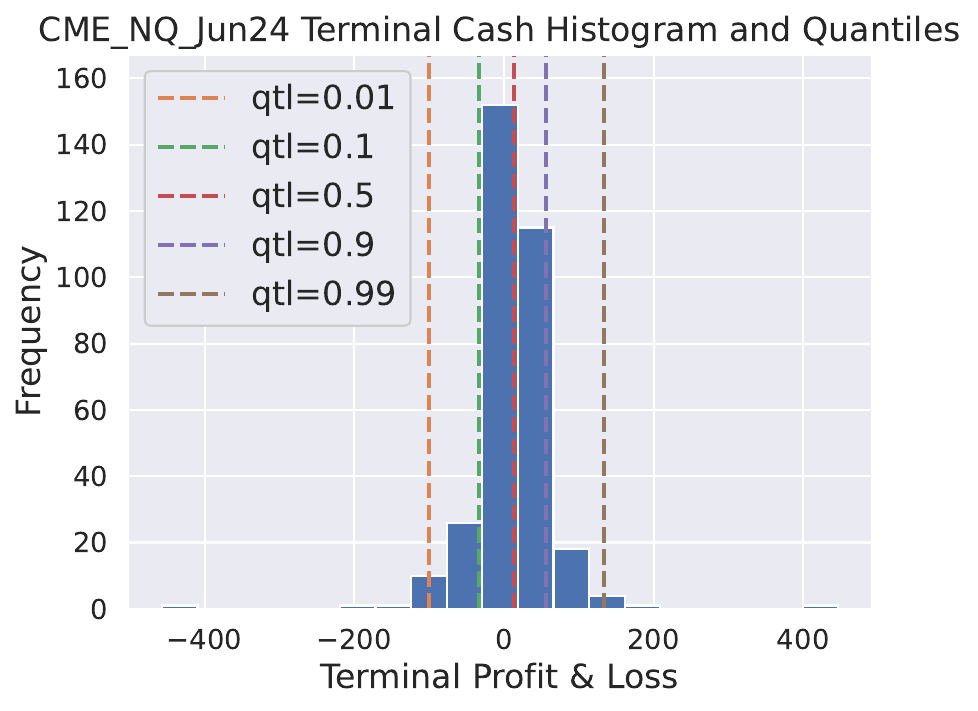}
        \end{subfigure}
        \hfill
        \begin{subfigure}[b]{0.4\textwidth}   
            \centering 
            \includegraphics[width=\textwidth]{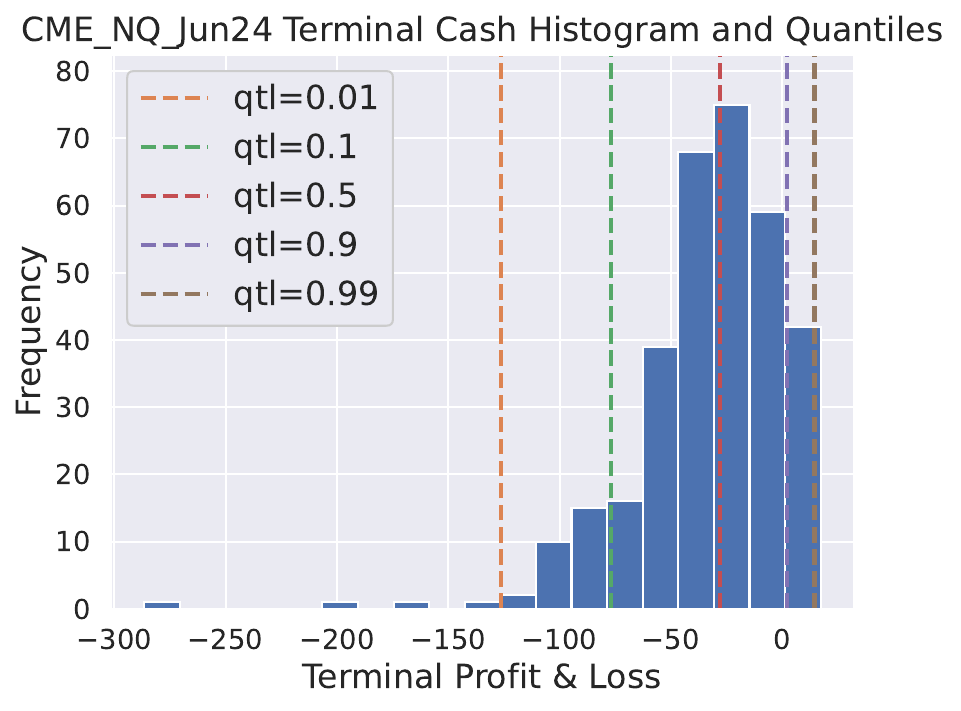}
        \end{subfigure}
        \vskip\baselineskip
        \begin{subfigure}[b]{0.4\textwidth}   
            \centering 
            \includegraphics[width=\textwidth]{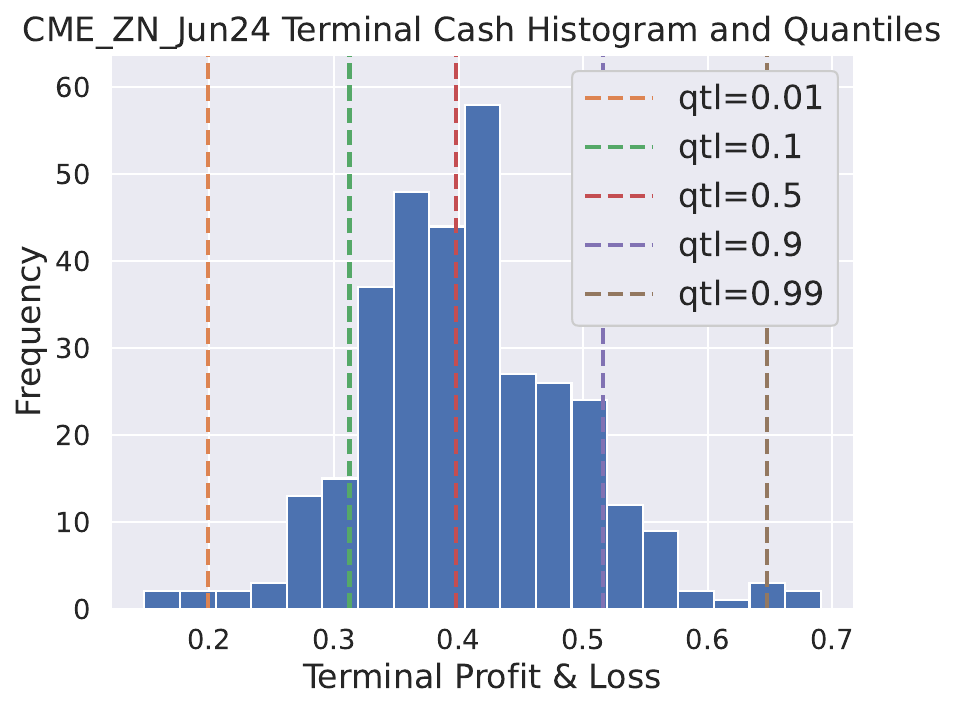}
        \end{subfigure}
        \hfill
        \begin{subfigure}[b]{0.4\textwidth}   
            \centering 
            \includegraphics[width=\textwidth]{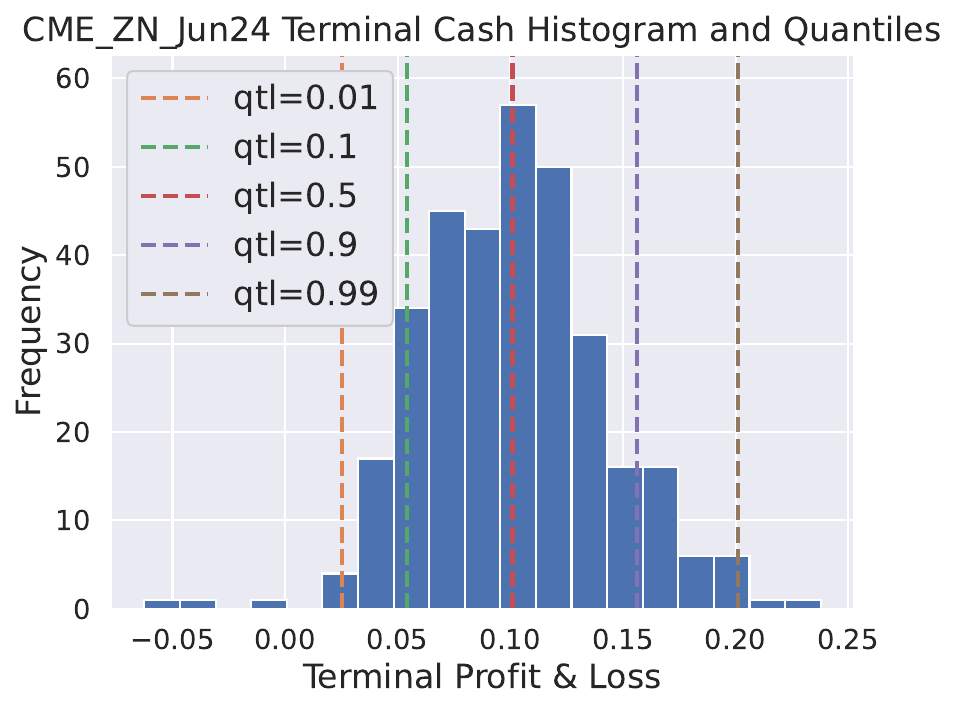}
        \end{subfigure}
        \captionsetup{font=small}
        \caption{ Terminal cash histogram in the benchmark (left) and improved (right) simulation environments for all 330 simulation paths. }
    \end{figure}

\begin{table}[H]
  \begin{center}
    
    \begin{tabular}{l|c|c|c|c|c|r} 
      \textbf{Fill Type} & \textbf{Amount} & \textbf{Fill Type} & \textbf{Amount} & \textbf{Fill Type} & \textbf{Amount}\\
      \hline
      AFA & 5804 & AFA & 11509 & AFA & 389\\
      NFA & 2107 & NFA & 1963 & NFA & 1960\\
      AFB & 5780 & AFB & 11448 & AFB & 432\\
      NFB & 2088 & NFB & 2009 & NFB & 1999
    \end{tabular}
    \captionsetup{font=small}
    \caption{Number of fills received over each 120 second run of the strategy that were adverse and non-adverse in ES, NQ and ZN, from left to right.  }
  \end{center}
\end{table}
Note that the non-adverse fill probability should be much lower in ZN as queue sizes are much larger relative to CL, ES, and NQ while trade sizes are very similar, hence the different distribution and more inflated performance results. 

\end{document}